\documentclass[nonacm,acmsmall,screen,letterpaper]{acmart}

\usepackage{subcaption}
\usepackage{tabularx}
\usepackage{xspace}
\usepackage{amsmath}
\usepackage{amsfonts}

\usepackage{tikz}
\usetikzlibrary{positioning}
\usepackage{amsmath}
\usepackage{amsfonts}
\usepackage{pdflscape}
\usepackage{bbm} 

\newcolumntype{s}{>{\hsize=0.30\hsize}X}
\usepackage{csquotes}

\newcolumntype{t}{>{\hsize=0.8\hsize}X}

\AtBeginDocument{%
  \providecommand\BibTeX{{%
    \normalfont B\kern-0.5em{\scshape i\kern-0.25em b}\kern-0.8em\TeX}}}

\setcopyright{acmcopyright}
\copyrightyear{2023}
\acmYear{2023}

\begin{document}

\title[Twits, Toxic Tweets, and Tribal Tendencies]{Twits, Toxic Tweets, and Tribal Tendencies: Trends in Politically Polarized Posts on Twitter}

\author{Hans W. A. Hanley}
\email{hhanley@cs.stanford.edu}
\affiliation{%
  \institution{Stanford University}
  \streetaddress{450 Serra Mall}
  \city{Stanford}
  \state{California}
  \country{USA}
  \postcode{94305}
}


\author{Zakir Durumeric}
\email{zakir@cs.stanford.edu}
\affiliation{%
  \institution{Stanford University}
  \streetaddress{450 Serra Mall}
  \city{Stanford}
  \state{California}
  \country{USA}
  \postcode{94305}
}



\begin{abstract}
Social media platforms are often blamed for exacerbating political polarization and worsening public dialogue. Many claim that hyperpartisan users post pernicious content, slanted to their political views, inciting contentious and toxic conversations. However, what factors are actually associated with increased online toxicity and negative interactions? In this work, we explore the role that partisanship and affective polarization play in contributing to toxicity both on an individual user level and a topic level on Twitter/X. To do this, we train and open-source a DeBERTa-based toxicity detector with a contrastive objective that outperforms the Google Jigsaw Perspective Toxicity detector on the Civil Comments test dataset. Then, after collecting 89.6~million tweets from 43,151~Twitter/X users, we determine how several account-level characteristics, including partisanship along the US left-right political spectrum and account age, predict how often users post toxic content. Fitting a Generalized Additive Model to our data, we find that the diversity of views and the toxicity of the other accounts with which that user engages has a more marked effect on their own toxicity. Namely, toxic comments are correlated with users who engage with a wider array of political views. Performing topic analysis on the toxic content posted by these accounts using the large language model MPNet and a version of the DP-Means clustering algorithm, we find similar behavior across 5,288~individual topics, with users becoming more toxic as they engage with a wider diversity of politically charged topics. 

\end{abstract}



\maketitle

\section{Introduction}
\vspace{1pt}
\noindent\fbox{%
    \parbox{.99\columnwidth}{%
        \textbf{Content Warning}: This paper studies online toxicity. When
        necessary for clarity, this paper quotes user content
        that contains profane, politically inflammatory, and hateful content.
    }%
}
\vspace{1pt}

\noindent
Over the past decade, political polarization within the United States has increased substantially~\cite{hong2016political,chen2022misleading,gaughan2016illiberal,gervais2015incivility,borah2013interactions,goovaerts2020uncivil}. Many people attribute the increase in division to social media, arguing that social media creates toxic political echo chambers where users become more politically polarized, reinforcing their biases~\cite{sunstein2018social,wojcieszak2022most}. In several documented cases, political polarization and associated toxicity have negatively impacted platforms, online communities, and users, sometimes leading to users leaving platforms altogether~\cite{pew-2017}. While many studies have investigated the role that toxicity and political polarization have had on the health of online communities ~\cite{tucker2017liberation,tucker2018social,torres2022manufacture,persily20172016,gron2020party,saveski2021structure}, there has been little work that investigates the role of toxicity, partisanship, and affective polarization (\textit{i.e.}, the tendency to be negative to those with different political views and positive to those with similar political views) between individuals and at the topic-level, the common means by which conversations take place on Twitter/X across multiple Twitter threads~\cite{wieringa2018political, quercia2012social,arslan2022understanding}. To fully understand the intertwined relationship between toxicity, partisanship, and polarization, at the user and topic-level in this work, we investigate:

\begin{enumerate}
    \item \textit{What are the relationships between partisanship, political polarization, and the tendency for politically engaged users to post toxic content?}
    \item \textit{How do the characteristics of users, including their partisanship, predict the toxicity of topics on Twitter/X?} 
\end{enumerate}

To answer these questions, we collect 89.6M~tweets from 43.1K~accounts throughout 2022. From these tweets, we measure the number of toxic tweets and toxicity of each user by designing and deploying our own DeBERTa-based~\cite{he2022debertav3} toxicity detection model, finding that it outperforms Google Jigsaw API~\cite{perspectiveapi}, the gold-standard out-of-box classifier for identifying uncivil and toxic language (\textit{e.g.}, insults, sexual harassment, and threats of violence~\cite{thomas2021sok}). Then, calculating each user's approximate political orientation using Correspondence Analysis~\cite{barbera2015tweeting} and performing fine-grained topic analysis using a large language model, we subsequently determine the interconnection between toxicity and political polarization at a user and topic-level.

\vspace{2pt}\noindent
\noindent
\textbf{RQ1: User-Level Factors of Toxicity and the Role of Political Polarization.} We first determine, using a linear regression model, some of the most significant features that predict the toxicity of content posted by individual Twitter accounts. We find that the most important feature that predicts an individual account's toxicity is the toxicity of the other accounts with which the user interacts. Namely, as users interact with other users who regularly tweet in a toxic manner, they themselves are more likely to tweet toxic content. We further find that while the position that a user falls on the political spectrum \emph{does not} have much bearing on the toxicity of their messages, the more that a given user interacts with users of different political orientations, the more likely their posts are to be toxic.  

\vspace{2pt}\noindent
\noindent
\textbf{RQ2: Toxicity and Political Polarization in Toxic Topics.} Having observed that users who interact with users of differing political views are more likely to be toxic, we examine this dynamic at a topic-level. After identifying 5.5M~English-language toxic tweets, we perform topic analysis using a fine-tuned version of the large language model MPNet and the DP-Means clustering algorithm~\cite{hanley2023partial}. Examining these topic clusters, we find that, in aggregate, the political orientation of users tweeting about a topic does not have a large effect on the topic's overall toxicity; rather we find that the effect of the political orientation of the users tweeting about particular topics varied widely. Examining factors that predict each topic cluster's overall toxicity, we find, as largely expected, that high-toxicity topics often involve high-toxicity users.  We further find that as individuals participate in a wider range of political topics the toxicity of their tweets increases. Namely, we identify at the topic-level (as on a user-level), a strong tribal tendency/affective polarization, with accounts acting negatively toward accounts of differing views. 

\vspace{3pt}
\noindent
Altogether, our work illustrates that, across a diverse set of users and topics, as engagement with toxic content and with a wider range of political views increases, so does average toxicity. In addition to open-sourcing a new toxicity classifier that achieves better accuracy than the Perspective API on the Civil Comments dataset, our work, one of the first to perform this analysis on a large-scale dataset of politically engaged users and across multi-thread topics not directly chosen by specific hashtags, illustrates how political polarization can negatively affect online communities and lead to increased divisiveness, regardless of the topic. We hope that this work helps inform future research into the role of polarization and toxic content in negatively affecting the health of online communities and intra-platform user interactions. 
\section{Background \& Related Work} In this section, we detail several key definitions utilized within our study, provide background on Twitter, and finally present an overview of existing works that inform our study.

\subsection{Terminology}\label{sec:misinformation-defintion}
We first provide some preliminary definitions of terms that form the basis of this work:

\vspace{2pt}
\noindent
\textbf{Online Toxicity and Incivility:} We utilize the Perspective API's definition of online toxicity and incivility: ``\textit{(explicit) rudeness, disrespect or unreasonableness of a comment that is likely to make one leave the
discussion.}'' given its extensive use in past studies of online toxicity~\cite{hua2020characterizing, saveski2021structure, xia2020exploring, kumar2023understanding}.

\vspace{2pt}
\noindent
\textbf{Political Partisanship:}  As in Barbera {et~al.}~\cite{barbera2015birds} and other works~\cite{saveski2022perspective,saveski2022engaging}, we define US political partisanship along a unidimensional axis ranging from left-leaning (\textit{i.e.}, liberal) to right-leaning (\textit{i.e.}, conservative). While this limits our analysis, given the variety of political views within the US, as found by Poole and Rosenthal, most of the variation in US political ideology \emph{is} along a unidimensional axis~\cite{poole2007party}, and this assumption is fairly common in the literature.

\vspace{2pt}
\noindent
\textbf{Affective Polarization:} 
Affective polarization is the tendency of individuals to distrust and be negative to those of different political beliefs while being positive towards people of similar political views~\cite{druckman2021affective}.

\subsection{Twitter/X}
Twitter is a microblogging website where users can post messages known as Tweets: messages with at most 280~characters. Tweets themselves, while often just text,  can also include hyperlinks, videos, and other types of media~\cite{jungherr2014twitter}. 
Unless made private, tweets are publicly displayed on the Twitter platform, allowing anyone to see or reply to the message~\cite{karami2020twitter}. 
As of late 2022, Twitter had approximately 238~million active daily users~\cite{Dang2022}. Many Twitter users get their daily news from the Twitter platform~\cite{boukes2019social,tandoc2016most,an2011media}. Despite the ability of anyone to gain and maintain a following on Twitter, several studies have found that political conversations are often dominated and guided by legacy media elites and celebrities~\cite{dagoula2019mapping}. We note that Twitter changed its name to X in mid-2023~\cite{Ivanova2023}, but for simplicity, we still refer to the platform as Twitter throughout this work.

\subsection{Political Partisanship and Polarization Online}
Various works have explored the role that individual users' political orientations play in interactions online. People, on the Internet and in their everyday interactions, tend to associate with like-minded individuals and Twitter is no exception~\cite{kamin2019social,huckfeldt1995political,halberstam2016homophily,barbera2014social,barbera2015tweeting,quattrociocchi2011opinions}. Several works have found that social media exacerbates this human tendency by creating political echo-chambers~\cite{starbird2018ecosystem}, where users' biases are reconfirmed and reinforced~\cite{conover2011predicting,cinelli2020echo,an2014partisan,bessi2016users}. Sunstein, Garett {et~al.}, and Quattrociocchi {et~al.} all argue that the ``individualized'' experience offered by social media companies comes with the risk of creating ``information cocoons'' and ``echo chambers'' that accelerate polarization~\cite{sunstein2018social,garrett2009echo,quattrociocchi2016echo}. Wojcieszak {et~al.}~\cite{wojcieszak2009online} determine that the majority of political discussions online are between participants who share the same viewpoint. Indeed, while the vast majority of Twitter users do not engage in political discussions, those that do, are often highly politically polarized~\cite{wojcieszak2022most}. 

As found by  Munson et~al.~\cite{munson2010presenting}, while some individuals seek views that are vastly different than their own, many also largely seek only affirming beliefs. Rogowski  et~al.~\cite{rogowski2016ideology} show that high ideological differences between individuals can lead to increased affective polarization; namely, if individuals are exposed to others with widely different beliefs, they increase their tendency to be negative toward those individuals and positive toward those who share their beliefs. Even more so, several recent research papers have found that social media can increase this rate of affective polarization~\cite{suhay2018polarizing,kubin2021role}. Cho et al.~\cite{cho2020search} find that exposure to social media content that attacks political figures can increase affective polarization. Most similar to our work, Bail~et~al.\cite{bail2018exposure} show that exposure to different political beliefs online can increase polarization, particularly for right-leaning individuals. 

In addition to polarization being amplified by social media, other works have found this increased polarization can increase misinformation and toxic behavior~\cite{an2014partisan}. Rains {et~al.}~\cite{rains2017incivility}, for instance, find that high polarization is a major factor in engendering online incivility and toxicity. Imhoff {et~al.}~\cite{imhoff2022conspiracy}, find that political polarization, on both sides of the political spectrum, is associated with beliefs in conspiracy theories. 

\subsection{Online Toxicity}

Online toxicity (\textit{e.g.}, doxing,  cyberstalking, coordinated bullying, and political incivility) plagues social media platforms~\cite{thomas2021sok,cuomo2019gender,kumar2021designing,nobata2016abusive,wulczyn2017ex,chandrasekharan2018internet}. As outlined by Thomas {et~al.}~\cite{thomas2021sok}, online toxicity is just one of type of hate and harassment, which intersects with other negative online behaviors like misinformation and extremism. Brubaker et~al.~\cite{brubaker2021power} find that trolls and bullies online are often motivated by a type of schadenfreude in spewing vitriol at other users. Similarly, Thomas et~al.~\cite{thomas2021sok} find that abusers are often also motivated by political ideology, disaffection, and control~\cite{thomas2021sok}. For example, a Flores-Saviaga~\cite{flores2018mobilizing} studied how users in the r/The\_Donald were motivated to troll and abuse other Reddit users in support of then-Republican candidate Donald Trump in 2016. In addition to harming the target, online toxicity often has many negative downstream effects. Kim {et~al.}, Kwon {et~al.}, and Shen {et~al.}, find, for example, that online toxicity is a self-reinforcing behavior, with negative conversations increasing observers' tendency to also engage in incivility~\cite{kim2019incivility,kwon2017offensive,shen2020viral}. Other works have found that marginalized groups often receive disproportionate amounts of toxicity online~\cite{relia2019race,thomas2021sok,chess2015conspiracy}. Pew Research, for instance, found that Black adults reported higher incidences of name-calling while women were more likely to experience sexual harassment. While toxicity can take many forms, in this work, we largely focus on toxic comments on Twitter.

\subsection{Detecting Online Toxicity }
Several works measure online toxicity using the Google Jigsaw Perspective API~\cite{perspectiveapi}. Saveski {et~al.}~\cite{saveski2021structure}, for example, utilize the Perspective API and find that many of the idiosyncrasies of particular Twitter conversations can lead to tweets with toxic language. Similarly, Habib {et~al.}~\cite{habib2022proactive}, utilize Perspective to identify opportunities for proactive interventions on Reddit before large escalations. Kumar {et~al.}~\cite{kumar2021designing} finally determine how different types of users interact with Reddit comments labeled by the Perspective API, finding that different social groups (\textit{e.g.}, women, racial minorities), often have different experiences when encountering the same comments. 

While the Perspective API has been utilized in a host of different recent studies~\cite{kumar2021designing,saveski2021structure,jain2018adversarial,rieder2021fabrics} likely because of its widespread adoption by large companies like Google, Disqus, Reddit~\cite{perspectiveapi}, several other works have sought to either improve on it utilizing newer large language models or non-machine-learning approaches. Grondahle et al.~\cite{grondahl2018all} show that adversarial training can make models robust to adversarial attacks like homoglyphs. Chandrasekharan et~al.~\cite{chandrasekharan2019crossmod} propose a cross-community learning strategy to build a model to help moderators on Reddit detect new context content. Lees et~al.~\cite{lees2022new} utilize a character-based transformer to build a state-of-the-art multilingual toxicity classifier that incorporates a learnable tokenizer allowing it to be robust to domains different from its training data. Kumar et~al. test recent large language models like GPT-4, Llama3, and Google Gemini, finding that they can account for ecosystems' norms and values when performing moderation~\cite{kumar2023understanding}. In contrast to these machine-learning approaches, Jhaver et~al.~\cite{jhaver2018online} illustrate the usefulness of the blocklists in better user experiences online. Finally, Lai et~al.~\cite{lai2022human} propose human-AI collaboration in detecting and removing content.

\subsection{Present Work}
Several works have studied how political polarization and online toxicity interact in particular political environments~\cite{cinelli2021dynamics,tucker2018social,bail2018exposure}. For example, Chen {et~al.}~\cite{chen2022misleading} utilize network analysis to find that misleading online videos lead to increased online incivility. Conversely, Rajadesingan {et~al.}~\cite{rajadesingan2021political}, find that political discussions in non-overtly political subreddits often lead to less toxic conversational outcomes. Most similar to our work, De~Francisci Morales {et~al.}~\cite{de2021no} find that the interaction of individuals of different political orientations increased negative conversational outcomes. In this work, however, rather than examining political polarization within a particular community or across one individual topic, we instead seek to understand across thousands of politically engaged users across the political spectrum, what are the most prominent characteristics that correspond with increased toxicity. Subsequently, our LLM-based approach, which identifies larger topic conversations across the tweets of politically engaged Twitter/X users and multiple Twitter threads~\cite{wieringa2018political, quercia2012social,arslan2022understanding}, then analyzes what contributes to polarized and toxic topics across political Twitter. Unlike previous approaches, which have largely relied on previously made hashtag lists, or were limited to a set of particular topics~\cite{cinelli2020echo} when analyzing the spread of topics, our approach is largely agnostic to these features, allowing us to analyze how various user and structural-level features contribute to toxicity across the Twitter platform. This thus approach enables us to study in a generalizable fashion how partisanship, and polarization, along with what characteristics contribute to negative and toxic outcomes across tweets about particular subjects of varying political salience.

\section{Methodology}
In this section, we provide an overview of how we collected our dataset and the algorithms that we utilize to understand the interactions among Twitter users and with different topics. 

\subsection{Estimating Partisanship\label{sec:background-ca}}

To approximate individual Twitter users' partisanship, we rely on the Correspondence Analysis (CA) proposed by Barber{\'a} {et~al.}~\cite{barbera2015tweeting}.  Correspondence analysis (CA), similar to principal component analysis, is a technique for categorical data that extracts discriminating and representative features from a given matrix~\cite{greenacre2010correspondence}. As found by Barber{\'a} {et~al.}, individual users often reveal their political preferences by whom they choose to follow on Twitter, and by analyzing these choices using CA, we can approximate their place on the political-ideological spectrum. CA works as follows: Given an $n \times m$ adjacency matrix that indicates whether user $i$ (row) follows user $j$ (column), CA can determine a discriminating latent space among these users based on their following behaviors. By carefully choosing our set of ``followed'' users (columns of the matrix) as a set of key political figures, this latent space can be used to represent a dimension of ``partisanship.'' Then, considering individuals' place on the left/right US political spectrum as a point within this latent space, we can estimate that point by projecting them onto the latent space based on who they choose to follow.\footnote{We utilize the Tweepy API to identify the set of users that each of our non-target political accounts follows.} The result is that if a given user follows many liberal-leaning/democratic or a set of accounts that liberal-leaning accounts tend to follow, then we consider that account to be liberal, and vice versa~\cite{barbera2015tweeting,mcpherson2001birds}. We note that with the CA technique, by later extending the set of the key followed accounts, this approach can be used to approximate the partisanship of users who do not necessarily follow one of the initial set of key political figures (\textit{e.g.}, congressional leaders).

\begin{figure}
\begin{minipage}[l]{0.47\textwidth}
\includegraphics[width=1\columnwidth]{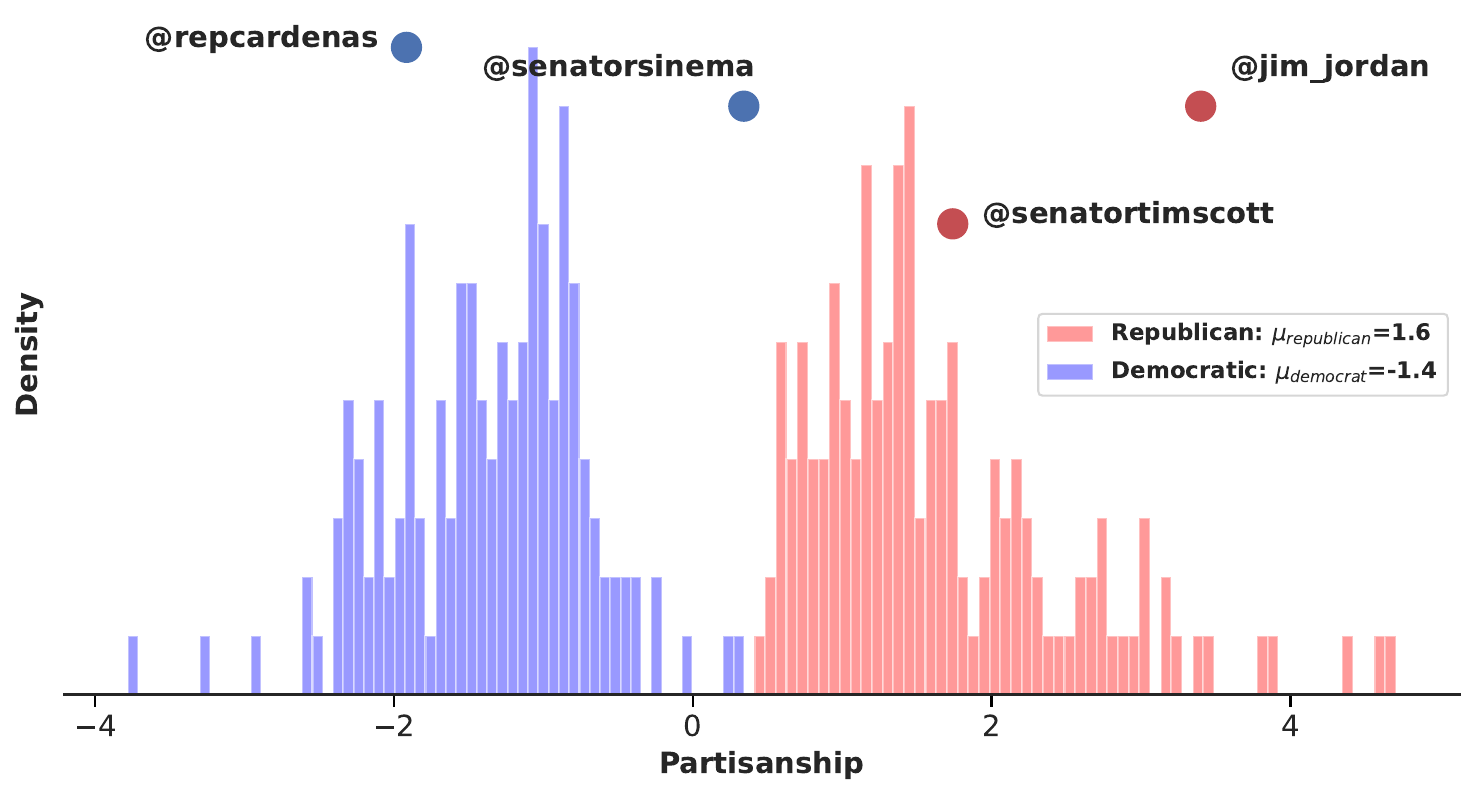} 
\end{minipage}
\begin{minipage}[l]{0.47\textwidth}
\includegraphics[width=1\columnwidth]{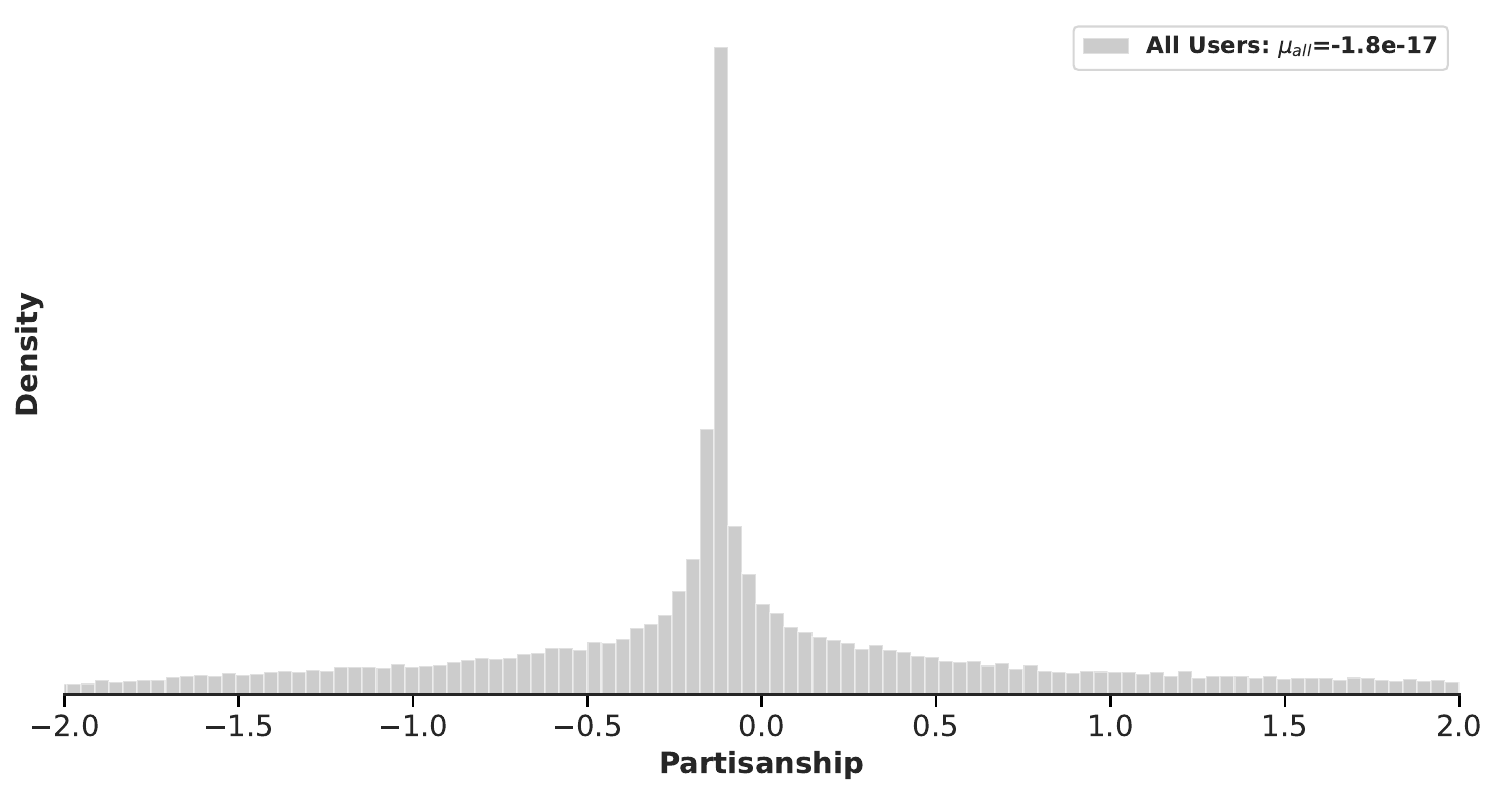} 
\end{minipage}

\begin{minipage}[l]{1\textwidth}
\caption{{Estimated Political Orientation of Political Leaders and All Users Using CA}-- We differentiate users' political leanings based on who they follow on Twitter.\label{fig:politican-orientation}}
\end{minipage}

\end{figure}
We note that for our initial set of key political predictive ``followed'' accounts,  we utilize the Twitter accounts of the US House of Representatives and US Senate members from the 117th Congress (2021--2023). In addition to these accounts, we further add another 352~political accounts that were formerly identified by Barber{\'a} {et~al.} (\textit{e.g.,} $@$JoeBiden, $@$VP).\footnote{\url{https://github.com/pablobarbera/twitter_ideology}} Using these accounts, and following the approach as specified by  Barber{\'a}  {et~al.}, we subsequently identified a politically ideological subspace and projected our final list of 43,151 different accounts to this subspace. See Appendix~\ref{sec:appendix-ca} for 
 additional details. As seen in Figure~\ref{fig:politican-orientation}, using this method we manage to obtain a discriminating latent space that allows us to differentiate the ideology of Republican and Democratic political leaders as well as our set of 43,151 accounts. In this setup, the more positive a user's ideology, the more right-leaning; the more negative, the more left-leaning.  

\subsection{Collecting Tweets}
Our dataset initially consisted of 187.6M tweets from 55.4K users that followed our set of key political figures. We collected this data utilizing the Twitter API throughout 2022. We note that following the acquisition of Twitter by Elon Musk, access to the API became restricted limiting our analysis to this time period~\cite{Singh2023}.  Upon identifying these users, given that our work is primarily focused on the US political system, we removed any user that listed their location on their Twitter profile as outside of the United States. To identify US-based users, we utilize the capability of the \texttt{Nominatim} Python tool to geo-code all user's locations based on their Twitter-provided location string and OpenStreetMap.\footnote{\url{https://www.openstreetmap.org/}} Altogether, we remove 12,264 users, leaving us 43,151 users. Upon identifying our user subset, we subsequently utilize \texttt{whatlango}\footnote{\url{https://github.com/abadojack/whatlanggo}} Go library to remove any non-English tweets from our set of users, leaving us 89,599,787 tweets. While we acknowledge several of our users' tweets might have been deleted or taken down by Twitter administrators before we scraped them, this dataset, consisting of over 89.6 million tweets, with an average of 2076.4 (median 614.0) tweets per individual is largely comprehensive of each user's tweeting behavior on the platform.

\subsection{Determining the Toxicity of Tweets\label{sec:labeling-toxic}}

We design and open-source\footnote{The weights for our model can be downloaded at \url{https://www.github.com/REDACTED}} a contrastive DeBERTa-based~\cite{he2022debertav3} model to determine the toxicity of tweets, later benchmarking our approach on two public datasets and against the Perspective Toxicity API~\cite{perspectiveapi}, the gold standard of toxicity detection~\cite{perspectiveapi,kumar2021designing,rajadesingan2020quick}. We note that throughout our work, we reproduce several results using the Perspective Toxicity classifier and present them in the appendix after obtaining similar results.
To train our new model we rely on the Civil Comments dataset\footnote{\url{https://www.kaggle.com/competitions/jigsaw-unintended-bias-in-toxicity-classification/data}} that was also utilized to train and validate the Perspective API. In addition to utilizing this dataset to augment our trained model, we take two main approaches: (1) data augmentation through realistic adversarial perturbations of the original Civil Comments dataset~\cite{le2022perturbations}, and (2) the inclusion of a contrastive learning embedding layer to help better differentiate toxic and non-toxic texts. For training details of our new model, see Appendix~\ref{sec:app-tox-classifier}.

\vspace{2pt}\noindent
\noindent
\textbf{Benchmarking our Toxicity Classifier.}
Upon training our toxicity model, we compare its performance against the Perspective Toxicity API~\cite{perspectiveapi} and a vanilla finetuned DeBERTa model with a classification head (a two-layer MLP with ReLU activation). To benchmark our toxicity model, we utilize the validation and test dataset of the Civil Comments dataset provided by Google Jigsaw\cite{perspectiveapi} as well as a separate toxicity dataset provided by Kumar {et~al.}~\cite{kumar2021designing}. Kumar  {et~al.}'s datasets consist of 107,620~social media comments (including from Twitter) where each comment was labeled by 5 human annotators as toxic or not (as opposed to the 10 annotators in the Civil Comments dataset). For our $F_1$ score calculations, as in Kumar {et~al.}~\cite{kumar2021designing} and in the Civil Comments dataset, we consider a comment to be toxic if its toxicity $t_i > 0.5$. Again, we utilize this threshold for classifying a comment as toxic, given that this score (as described in the Civil Comments task) indicates that a majority of the Civil Comments annotators would have assigned a ``toxic'' attribute to this comment.

As seen in Table~\ref{tab:benchmark-toxicity}, our contrastive DeBERTa model achieves the lowest mean absolute error (MAE) as well as the highest Pearson correlation and $F_1$ scores across the Civil Comments validation and test dataset. In addition, while obtaining a slightly lower correlation, our model on this separate dataset achieves a lower mean absolute error and a higher $F_1$ score. As such for the rest of this work, when determining the toxicity of tweets, we utilize our contrastive DeBERTa model. We note that our model has a $\rho= 0.870$ Pearson correlation with the scores output by the Perspective API, illustrating its use as an offline alternative with competitive performance to Perspective. Lastly, for this work, as in other works~\cite{hanley2023sub, rajadesingan2020quick}, when determining the overall toxicity of users, or particular groupings of tweets, we utilize the average of the toxicity scores of the tweets output by our model.

\begin{table*}
\centering
\fontsize{8.4pt}{6}
\selectfont
\begin{tabular}{l|ccc|ccc|ccc}
\toprule
& \multicolumn{3}{c|}{\textbf{CC Validation}} & \multicolumn{3}{c|}{\textbf{CC Test }} & \multicolumn{3}{c}{\textbf{Kumar {et~al.} }} \\

Model & MAE & Corr. &Macro-$F_1$ & MAE & Corr. &Macro-$F_1$&  MAE & Corr. &Macro-$F_1$\\ \midrule
DeBERTa & 0.0650 &0.800 &0.841 & 0.0654& 0.797 &0.842 & \textbf{0.241} & 0.383 & 0.539  \\
DeBERTa-contrastive & \textbf{0.0601} &\textbf{0.820} &\textbf{0.851} & \textbf{0.0609}& \textbf{0.818}&\textbf{0.852} & {0.251} & 0.415 & \textbf{0.540}  \\
Perspective API & 0.0961& 0.778& 0.845& 0.0963& 0.777 & 0.842 & 0.332 & \textbf{0.417} & 0.410 \\
\bottomrule
\end{tabular}
\vspace{3pt}
\caption{\label{tab:benchmark-toxicity} Mean absolute error, Pearson correlation, and $F_1$ score of the Perspective API and our DeBERTa models on the Civil Comments Validation and Test dataset. We bold the best scores in each respective column  }
\vspace{-10pt}
\end{table*}

\subsection{Topic Analysis with MPNet and DPMeans\label{sec:topic-background}} To later understand how particular types of users interact with different topics composed of toxic tweets, we perform topic analysis on these messages. As found by Grootendorst {et~al.}~\cite{grootendorst2020bertopic,hanley2023partial}, by embedding small messages like Tweets into a shared embedding space and then clustering these embeddings, fine-grained and highly specific topics can be extracted from datasets. To do this, we utilize the large language model MPNet\footnote{\url{https://huggingface.co/sentence-transformers/all-mpnet-base-v2}} fine-tuned on semantic search and a parallelizable minibatch version of the DP-Means algorithm.\footnote{\url{https://github.com/BGU-CS-VIL/pdc-dp-means}}

\vspace{2pt}\noindent
\noindent
\textbf{Fine-tuning MPNet for Topic Analysis.} To compare two tweets' semantic content for later clustering, we rely on a version of the MPNet~\cite{song2020mpnet} large language model that was fine-tuned on semantic search. MPNet maps sentences and paragraphs to a 768-dimensional space, comparing different sentence and paragraph embeddings' semantic content based on cosine similarities (ranging from -1 [highly different] to +1 [highly similar]). We note that the version of MPNet that we utilize was initially fine-tuned on similar social media data (\textit{e.g.}, Reddit comments, and Quora Answers) allowing us to apply this model to our set of tweets. However, to further ensure that our MPNet model is properly suited to our Twitter dataset, as in Hanley et al.~\cite{hhanleyspecious2024}, we further fine-tune this model using an unsupervised contrastive learning objective(\textit{i.e.}, the SimCSE training objective) to better the quality of our embeddings~\cite{gao-etal-2021-simcse}  on our set of tweets. As training data for this fine-tuning, we utilize 1 million tweets randomly sampled from our set of 89.6 million tweets.  See Appendix~\ref{sec:finetune} for additional details. As a reference, we provide two example tweet pairs with similarities at 0.55 and -0.18 in Figure~\ref{figure:paragraph_pairs}. We note that for each tweet within our dataset, before embedding the message, we first remove all URLs, ``@'', ``\#'', emojis, photos, and other non-textual elements from the message. In addition, for each user handle or text hashtag that utilizes camel case (\textit{i.e.}, camelCase) or snake case (\textit{i.e.}, snake\_case), we finally unroll those strings to their constituent elements.

\begin{figure}

\begin{minipage}{1\textwidth}
\hspace{100pt}
\tiny
\textbf{0.735 similarity}
\hspace{120pt}
\tiny
\textbf{-0.032 similarity}
\end{minipage}

\noindent\fcolorbox{black}{lightgray}{%
\begin{minipage}{.44\textwidth}
\tiny
\textbf{Tweet 1:} AZ voters; we need your vote for 
@Adrian\_Fontes and the rest of the Blue ticket. Please, for us, for our children. Please.
\textbf{Tweet 2:} Please vote for 
@Adrian\_Fontes
 and save AZ
\end{minipage}}
\noindent\fcolorbox{black}{lightgray}{%
\begin{minipage}{.44\textwidth}
\tiny
\textbf{Tweet 1:}If a supervisor was giving off those kinds of vibes to a worker in a conventional workplace, an HR complaint would definitely be warranted. Creepy AF

\textbf{Tweet 2:} The close relationship between politics and economics is neither neutral nor coincidental. Large governments evolve through history in order to protect large accumulations of property and wealth.”
\end{minipage}}
\caption{Examples of  Tweet pairs at different similarities (0.735 left and -0.032 right).  }

\label{figure:paragraph_pairs}
\vspace{-10pt}
\end{figure}

\vspace{2pt}\noindent
\noindent
\textbf{DPMeans for Clustering Tweets.}
DP-Means~\cite{dinari2022revisiting} is a non-parametric extension of the K-means clustering algorithm. When running DP-Means, when a given datapoint is a chosen parameter $\lambda$ away from the closest cluster, a new cluster is formed, and that datapoint is assigned to it. This characteristic of DP-Mans enables us to specify how similar individual items must be to one another to be part of the same cluster. Similarly, because DP-Means is non-parametric in terms of the number of clusters formed, we do not need to know \textit{a priori} how many topics are present within our dataset. For additional details about DP-Means, see Appendix~\ref{sec:ap-dpmeans}.

\begin{figure}

\includegraphics[width=0.9\columnwidth]{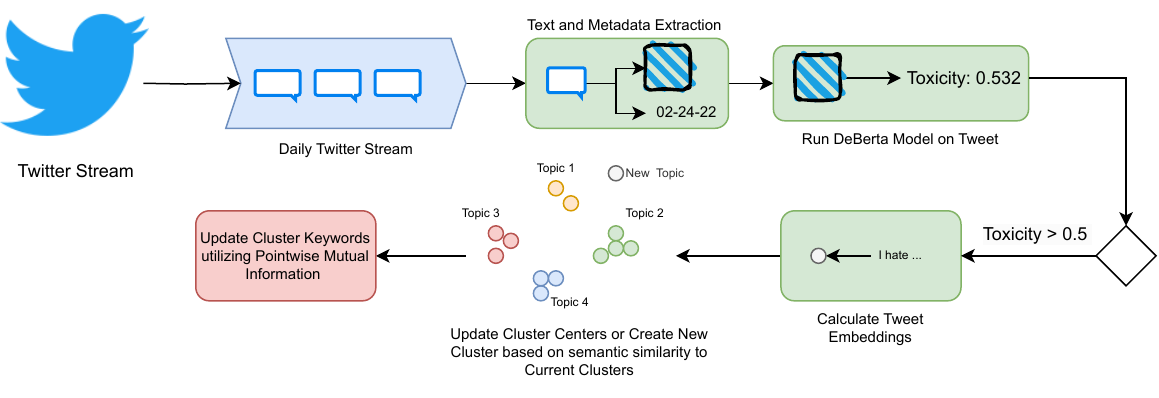} 
 \caption{Topic analysis of Toxic Tweets---We determine the toxicity, embed, and cluster toxic tweets to identify the most polarized and toxic conversations on Twitter throughout 2022. We note that for this approach, we limit our analysis to English tweets. We utilize the \texttt{whatlango} Go library to determine the language of tweets. \label{fig:toxic-cluster}}
\end{figure}

\vspace{2pt}\noindent
\noindent
\textbf{Topic Analysis Pipeline.}
Having outlined the constituent elements of our topic analysis algorithm, we now go over the full topic analysis pipeline (Figure~\ref{fig:toxic-cluster}): Throughout 2022, as we gathered the tweets of our set of 43,151 Twitter users, using our DeBERTa-contrastive model, we identify potentially toxic tweets (\textit{i.e.}, toxicity $t_i$ > 0.50). Following the identification of these potentially toxic tweets and separating out non-English tweets with \texttt{whatlango}, using MPNet, we subsequently map these tweets to a shared embedding space. Finally, we continuously cluster these tweets to identify topics amongst these toxic tweets using the DP-Means algorithm. To make these clusters that represent topics amongst our set of tweets, human-understand we employ two different approaches. First, we designate the tweets closest (\textit{i.e.}, with the largest cosine similarity) to the center of the cluster as the ``representative tweet'' of the cluster~\cite{grootendorst2020bertopic}. Second, we determine the most distinctive keywords of each cluster using pointwise mutual information~\cite{bouma2009normalized} (detailed in Appendix~\ref{sec:pmi}). In this way, after clustering our set of tweets, we can later extract the semantic meaning of the various clusters outputted. 

As recommended by Hanley {et~al.} we utilize a $\lambda$ of 0.60 for our clusters (precision near 0.989 for MPNet~\cite{hanley2022happenstance,grootendorst2020bertopic}). Finally, we extract keywords from these clusters using the pointwise mutual information metric and determine the most representative tweets by determining the tweet with the highest cosine similarity to the cluster center. Altogether, across the 5,509,042~English-language toxic tweets from our set of 43,151~Twitter users, we identified 5,288~clusters with at least 50~toxic tweets. 

\subsection{Generalized Additive Models}
Throughout this work, we utilize Generalized Additive Models (GAM)~\cite{hastie2017generalized} to determine the relationships between our variables of interest (\textit{e.g.}, user partisanship, and user toxicity). For GAMs, the relationship between the independent and dependent variables is not assumed to be linear but is rather estimated as a smooth regularized nonparametric function. Namely, given a dependent variable $Y$ and a set of $p$ independent variables  $X$, GAM's are estimated as:
\begin{equation}
    g(E(Y))=\alpha+s_1(x_1)+\cdots+s_p(x_p),
\end{equation} where $g()$ is a linking function that connects the expected value of the dependent variable $Y$ to values of functions $s_i()$ of independent independent variables in $X$. For example, when estimating probabilities the logit function is often utilized as with ordinary Generalized Linear Models~\cite{demaris1992logit}. The functions $s_i()$ represent smooth nonparametric functions of the variables in $X$ that are fully determined by the data in $X$ rather than by a parametric function. For GAMs, these $s_i()$ are estimated simultaneously, and the estimated value of $g(E(Y))$ is determined by implying adding up the values of the $s_i()$ functions. Throughout this work, we utilize the Python \texttt{Pygam} library to fit our regressions and utilize the Generalized Cross Validation Loss Criteria (GCV)~\cite{demaris1992logit} for estimating the $s_i()$ functions when fitting. The Generalized Cross-Validation Loss Criteria takes a LOOCV (Leave One Out Cross-Validation) approach to fitting smoothers on the data in X. 

Utilizing GAMS versus other more traditional models allows us  (1) to not assume linear relationships between our dependent and independent variables, and (2) to have better interpretability given that the partial contribution of a given variable $x_i$ to determining the value of the dependent variable Y is a function only of its corresponding function $s_i()$. 

\subsection{Ethical Considerations} 
Within this work, we largely focus on identifying large-scale trends in how different Twitter interact with one another. While we do calculate toxicity and polarization levels for individual users, we only display the names of verified public users or users with more than 500K followers, redacting the names of all other accounts. 
We lastly note that our Twitter data was largely collected before Elon Musk's private acquisition of Twitter on October 27, 2022, and all of our data was collected before the later restrictions placed on the collection of tweets on June 30, 2023.\footnote{\url{https://help.twitter.com/en/rules-and-policies/twitter-limits}}

\section{RQ1:  User-Level Factors in Toxicity on Twitter\label{sec:toxic_middle}}
Having provided background on our methodology and dataset, in this section, we discuss several of the user-level factors that coincide with and contribute to the toxicity on Twitter.

\subsection{Setup}
Here, we examine the role of several user-level factors in contributing to or affecting the rate at which individual users are toxic on Twitter. Specifically, we examine the following user characteristics in contributing to or mitigating the toxicity of individual users on Twitter: 

\begin{enumerate}

\item \emph{The verified status of the account}
\item \emph{The number of years the account has been active on Twitter}
\item \emph{The log of the number of the account's followers}
\item \emph{The log of the number of accounts the user follows}
\item \emph{The account's partisanship as determined by our Correspondence Analysis}
\item \emph{The estimated average toxicity of all users the account mentioned/@ed on Twitter (\textit{i.e.}, accounts that the user has interacted with)}
\item  \emph{The estimated average partisanship of the accounts the user mentioned/@ed}
\item \emph{The standard deviation of the partisanship of the accounts that the user mentioned (\textit{i.e.}, the range of political views with which the user interacts)}
\item \emph{The average value of the partisanship of all accounts the user mentioned/@ed}
\item \emph{The average difference in the partisanship of the account the given user mentioned/@ed and the users's partisanship}

\end{enumerate}

\noindent We fit these ten covariates against each of our account's average toxicity scores. As in past studies, we fit against the verified status, the age of the account, and the information about the activity of the accounts (\textit{e.g.}, the number of followers and the number of users followed)  to understand how general account characteristics that the Twitter API returns correspond with user toxicity~\cite{hua2020characterizing, chatzakou2017measuring}. As shown in prior work, the verification status, the number of years active, and levels of activity, depending on the context, can have differing effects on the adversarial nature and toxicity of Twitter accounts~\cite{chatzakou2017measuring,ribeiro2018characterizing}. Similarly, as shown in Saveski et al.~\cite{saveski2021structure} and Kraut et al~\cite{kraut2010dealing}  many individual-level characteristics are predictive of users' toxicity as it predicts their level of familiarity with a given platform and their tendency to break norms (\textit{e.g.}, post toxic content).  Thus, as a baseline, and to help ground our study, and determine how these account characteristics correlate with increased toxicity within the context of politically US-aligned account interactions, we include them in our model. In addition to these basic account attributes, we include information about each Twitter account's partisanship on the US left-right political spectrum as well as information about how that Twitter user interacts with other US politically aligned Twitter accounts~\cite{marozzo2018analyzing}. These variables' inclusion allows us to answer our research question about whether and how affective polarization and partisanship affect the toxicity of individual accounts~\cite{kubin2021role}.

\begin{figure}
\begin{minipage}[l]{1.0\textwidth}
\includegraphics[width=1\columnwidth]{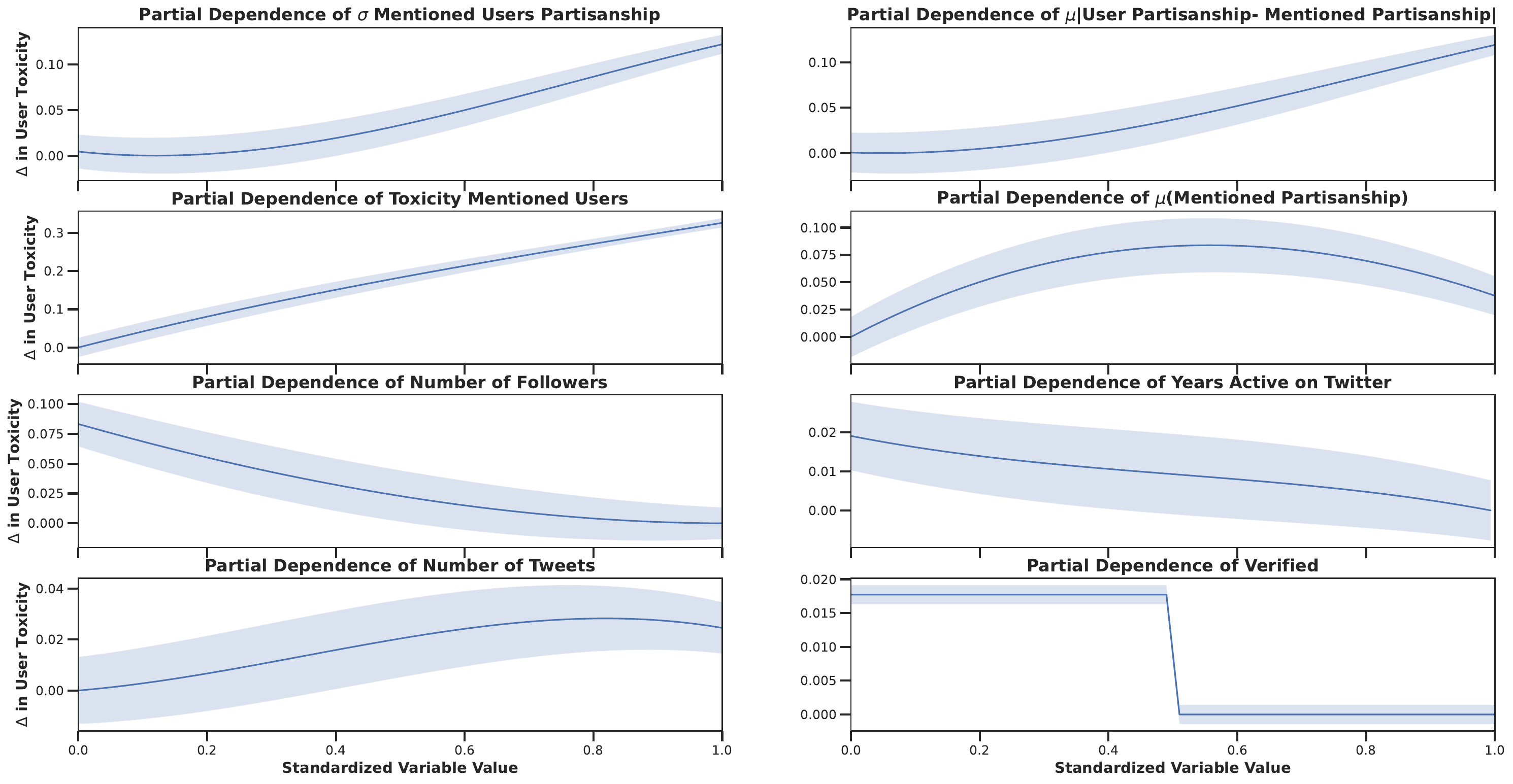} 
\end{minipage}
\begin{minipage}[l]{1\textwidth}
\caption{Partial dependencies with 95\% Normal Confidence intervals between our fitted standardized dependent variables and user toxicity.}
\label{fig:partial-dependence-user-toxcity}
\end{minipage}

\end{figure}

\begin{table}
    \small
    \centering
    \begin{tabularx}{0.90\columnwidth}{l|rrr}
    Train $R^2$:  0.239 ,   $R^2$:  0.207  \\
    \toprule
      Dependent Variable  & Pearson Corr. $\rho$ & Kendall's $\tau$  & Permut Import. \\    \midrule
  Verified Status & ---- &-0.242 & 0.053  \\
  Years Active on Twitter & -0.197& -0.147 & 0.027 \\
  Log \# Followers & -0.206 &-0.122 & 0.205  \\
  Log \# Followed & -0.135 & -0.090 & --- \\
  Log \# Tweets in 2022 & 0.147 & 0.200  & 0.045\\
  Toxicity of Mentioned Users& 0.318& 0.362 & 0.374 \\
  partisanship  &0.054& 0.061 & --- \\
  $\sigma$(Mentioned partisanship)&0.317& 0.332	& 0.150\\
   $\mu$(Mentioned partisanship) & 0.110 &0.099 & 0.067  \\
  $\mu$|User partisanship- Mentioned partisanship|&0.287& 0.283 & 0.080\\

    \bottomrule
    \end{tabularx}
  \caption{Pearson correlation $\rho$ and Kendall's $\tau$, and permutation importance of dependent variables and user's toxicity. As seen in the above table, a user's interaction with a wide political variety of users and interacting with other users with higher toxicity correlates with a given user's toxicity. } 
   \vspace{-15pt}
   \label{table:importance-user-toxicity}
\end{table}

To understand how these factors interact with and contribute to toxicity on Twitter, we fit a Generalized Additive Model (GAM) on the average toxicity score of users (Table~\ref{table:importance-user-toxicity}). When fitting our model, we perform variable selection using forward selection based on the Akaike Information Criterion~\cite{akaike2011akaike}, which ended up eliminating the number of followed accounts as well as the user partisanship as variables from our final model. Furthermore, to ensure that our model generalizes, we further reserve 10\% of our data as validation, and in our results report our model's $R^2$ value on this validation set. Finally, after fitting this regression, we further determine the estimated importance of each variable to our final model by permuting the features and seeing the estimated impact on the $R^2$ score on the validation set of our data (permutation importance is a widely used statistic for determining the relative information of features to models~\cite{altmann2010permutation}). We present the partial dependence (with 95\% Normal confidence intervals) on the user toxicity of each independent variable in Figure~\ref{fig:partial-dependence-user-toxcity} and present Pearson correlation, Kendall's $\tau$  (a more robust version of the Spearman Correlation), and each independent variable's permutation importance in Table~\ref{table:importance-user-toxicity}.  Our final model achieved a $R^2$ value of 0.239 on our training data and a $R^2$ value of 0.207 on our validation dataset, illustrating that our model does generalize to users outside of its training data.

We lastly note that to ensure the robustness of our approach, we separately perform the same analysis utilizing the toxicity scores output by the Perspective API, obtaining similar results. We present these results in Appendix~\ref{sec:perspective-user-app}.

\subsection{Baseline Account Characteristics}
We first provide an overview of how several baseline account characteristics contribute to the toxicity of each user. As seen in Table~\ref{table:importance-user-toxicity}, we do indeed observe that each of the user characteristics that we consider (to varying degrees) \emph{does} indeed have observed a correlational effect on how toxic users' tweets tend to be. We consider each of these effects below.

\vspace{2pt}\noindent
\noindent
\textbf{Verified Status.} As seen in Table~\ref{table:importance-user-toxicity} and Figure~\ref{fig:partial-dependence-user-toxcity}, as also found by Hua {et~al.}~\cite{hua2020characterizing}, whether a user is verified has a modest effect on how often they post toxic tweets, with verified users being less likely to tweet harmful or toxic messages compared to non-verified users. Overall, we find that a user's verification status has a Kendall's $\tau$ of -0.242 with their verification status and has a permutation importance of 0.053 in our final model. This suggests that when users become verified and their account is associated with their offline life, users tend to be less toxic. We note that we collected users' verification status before the implementation of Twitter Blue (users could pay 8 USD to become verified) in November 2022~\cite{Fung2023}.

\vspace{2pt}\noindent
\noindent
\textbf{Years Active on Twitter.} As users stay on Twitter, as seen in Figure~\ref{fig:partial-dependence-user-toxcity}, we observe that they are less likely to be toxic. As argued by Rajadesingan {et~al.}~\cite{rajadesingan2020quick} in their paper on Reddit, as social media users stay longer on particular platforms and adjust to interacting with other users, they tend to be less aggressive and toxic with other users. We see a similar result here, with older users being less toxic than younger ones. Overall, we observe that the number of years that a user is active on Twitter has a Pearson correlation of $\rho=-0.197$ with their average toxicity and a permutation importance of 0.027. This accords with past research that has found that news users, who are used to the social mores and norms of a given online community, may more frequently violate those norms and post toxic content~\cite{kraut2010dealing}.

\vspace{2pt}\noindent
\noindent
\textbf{Number of Followers.} Like verified status, and as argued by Marwick {et~al.}~\cite{marwick2011see}, extremely popular users are less likely overall to be toxic than users with smaller followings. These users, often create friendly public personas to interact with their followers, rarely attacking other users or posting toxic content. As seen in Figure~\ref{fig:partial-dependence-user-toxcity}, we see the same: more popular users that have more followers are less likely to post toxic tweets ($\rho=-0.206$). This variable has a permutation importance of 0.205 suggesting the high relative importance in determining the toxicity of accounts. 

\vspace{2pt}\noindent
\noindent
\textbf{Number of Tweets.}
Many accounts in our Twitter dataset post several times a day, with the median account posting 614.0~times throughout 2022, and one account posting 413,658 times. As seen in Figure~\ref{fig:partial-dependence-user-toxcity} with a permutation importance of 0.045 and a Pearson correlation of $\rho=0.147$, we observe that as Twitter users post more, generally their average toxicity increases. This finding reinforces past work that suggests that accounts that post excessively and that spam Twitter, are more likely to be toxic~\cite{salehabadi2022user}. 


\begin{figure}
\begin{minipage}[l]{0.6\textwidth}
\includegraphics[width=1\columnwidth]{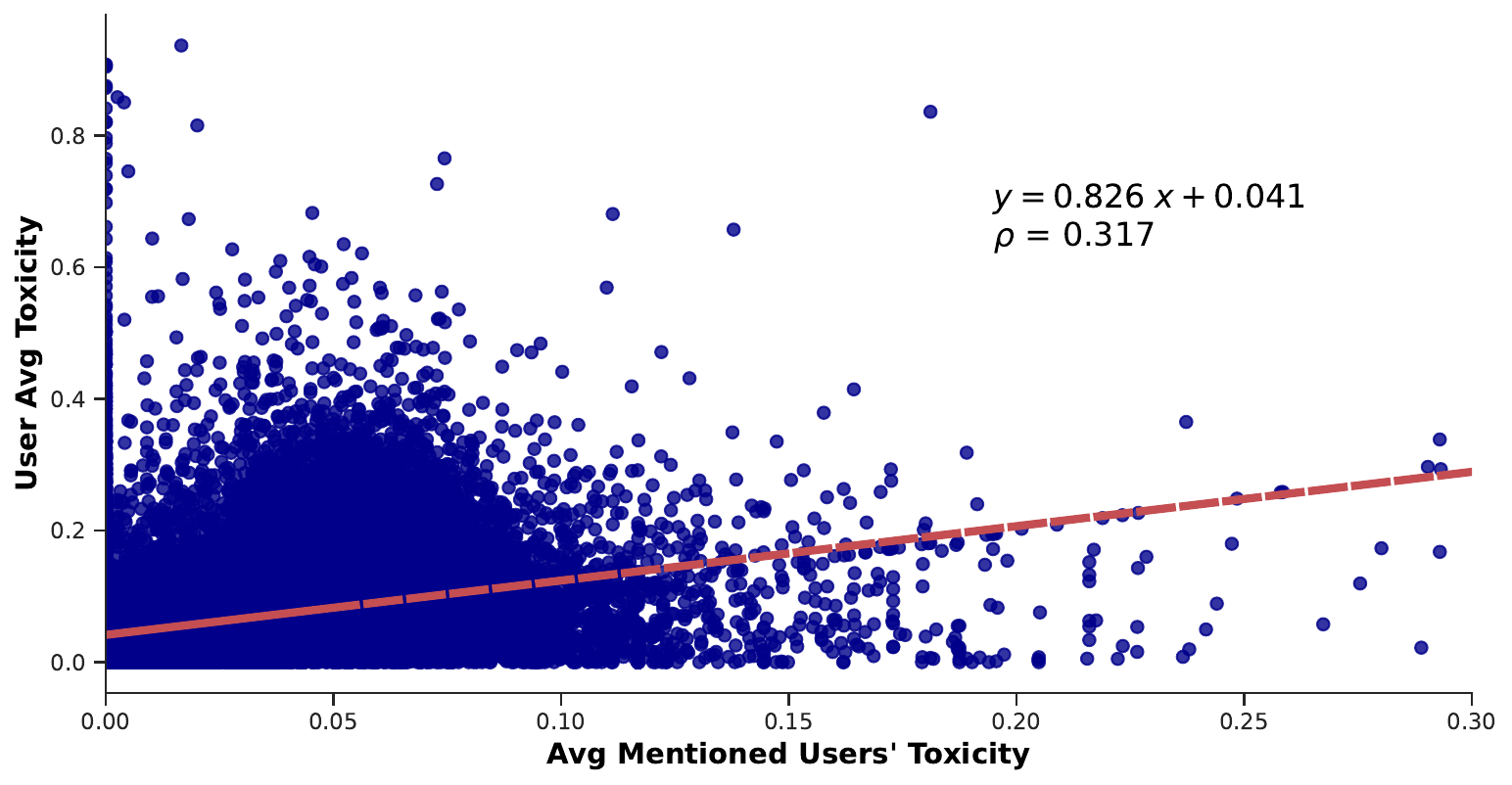} 
\end{minipage}
\begin{minipage}{.33\textwidth}
  \centering
  \includegraphics[width=1\linewidth]{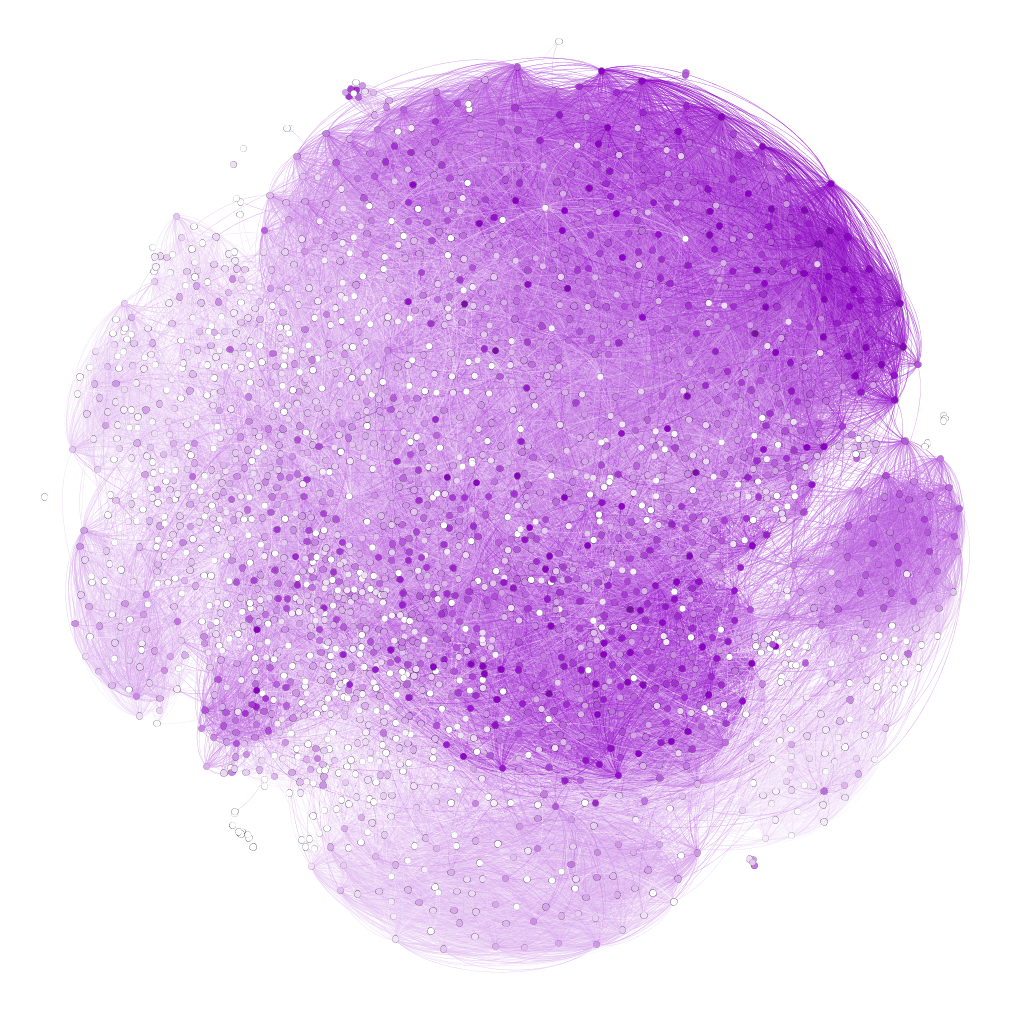}
  \label{fig:Twitter-reddit-toxicity-time}
\end{minipage}
\begin{minipage}[l]{1\textwidth}
\caption{The more toxic the users mentioned by a given user, on average, the more toxic the content of that particular user. Within the mention graph (the darker the purple the more toxic) of user interactions, toxicity has an assortativity coefficient of 0.071, suggesting that, to some degree, users who post toxic content have a slight tendency to mention and interact with other users who post toxic content. \label{fig:toxicity-vs-mention}}
\end{minipage}

\end{figure}

\subsection{Calculated Account Characteristics: Toxicity and Political Orientation\label{sec:calculated}}

Here we provide an overview of how the different political and toxicity measures that we calculated contribute to individual user-level toxicity. 

\vspace{2pt}\noindent
\noindent
\textbf{Toxicity of Mentioned Users.} We find that as users interact with or mention (@ing) other users who post toxic content, they themselves are more likely to be toxic. As seen in Figure~\ref{fig:partial-dependence-user-toxcity}, the average toxicity of accounts with which a user interacts has a nearly linear relationship with the user's own toxicity with very little variation. Indeed, we find this variable to be the most important in determining a user's toxicity, with it having a permutation importance of 0.374 and a Pearson correlation $\rho=0.318$. The most important of our covariates in terms of explainability, this result reinforces many prior findings about when and why particular users are toxic online~\cite{saveski2021structure,rajadesingan2020quick}. Creating a mention (@) graph among our 43,151~users and plotting users' toxicity against the toxicity of their mentioned accounts in Figure~\ref{fig:toxicity-vs-mention}, we further find some degree of assortativity based on toxicity (0.071), with more toxic users more likely to interact with each other than with non-toxic users, supporting this result.

\begin{figure}
\begin{minipage}{.33\textwidth}
  \centering
  \includegraphics[width=1\linewidth]{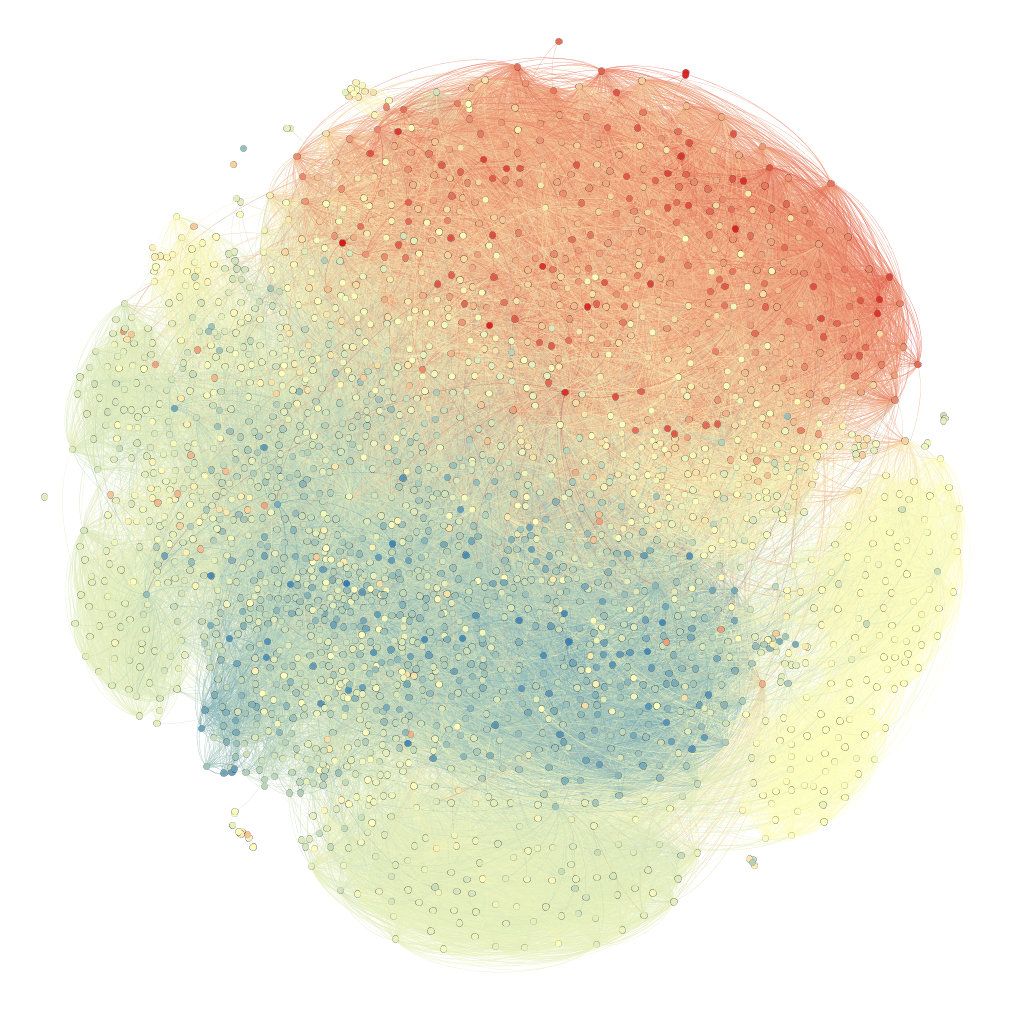}
\label{fig:twitter-reddit-partisanship-time}
\end{minipage}%
\begin{minipage}{.6\textwidth}
  \centering
  \includegraphics[width=1\linewidth]{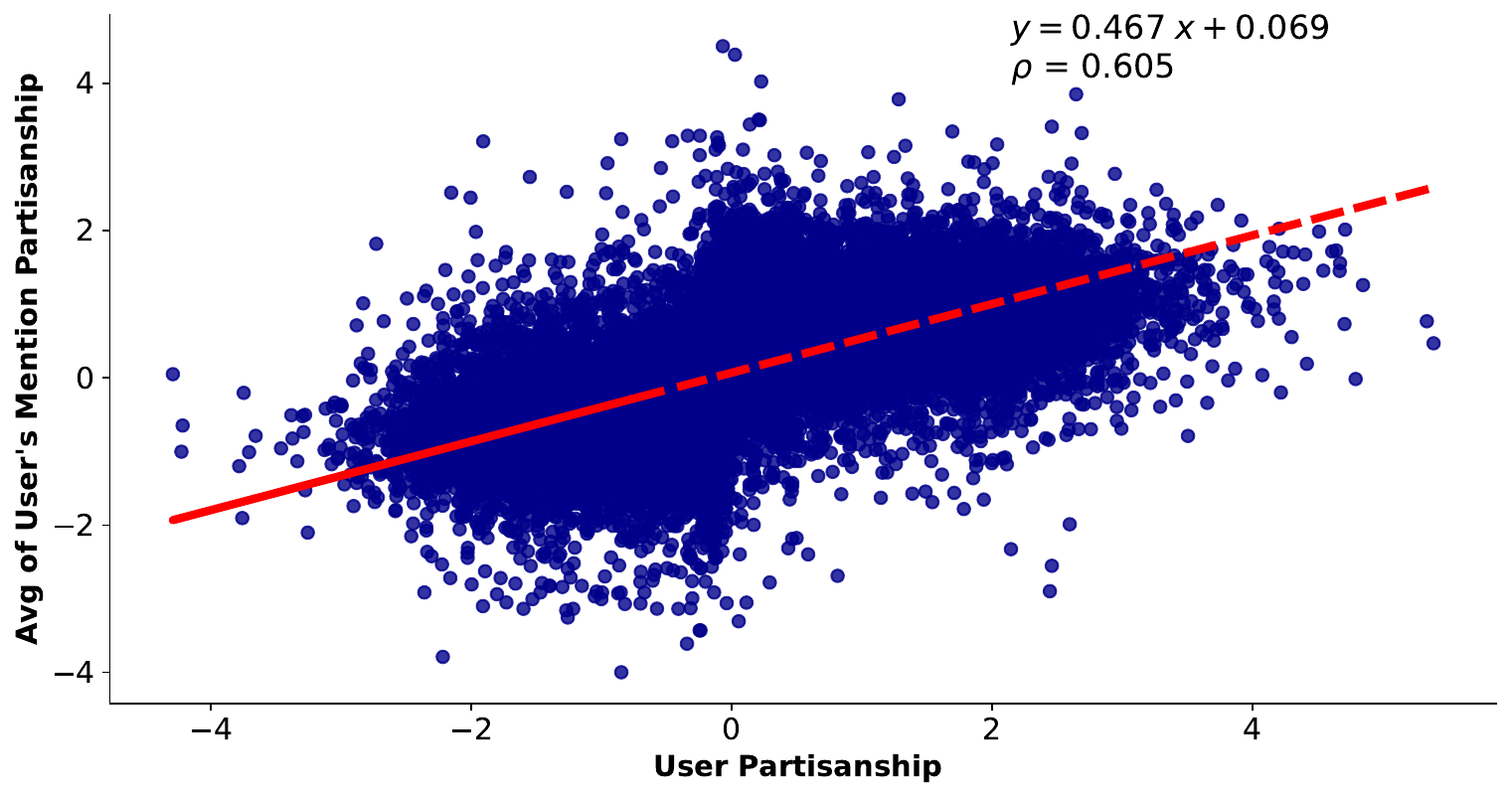}
\label{fig:twitter-reddit-partisanship-time2}
\end{minipage}%

\begin{minipage}{1\textwidth}
\caption{Within the mention graph of user interactions (red/right-leaning and blue/left-leaning), partisanship has an assortativity coefficient of 0.266, suggesting that conservative users mention and interact more with right-leaning users while liberal users interact more with and mention other left-leaning users. Similarly, graphing the average of each user's mention's partisanship against their own partisanship, we find significant assortativity (Pearson correlation $\rho=0.605$)\label{fig:graph-of-political-interactions}
}
\end{minipage}

\end{figure}
\vspace{2pt}\noindent
\noindent
\textbf{Partisanship of Mentioned Users.} As the average value of the partisanship increases (the mentioned accounts become more right-wing), we find that the average toxicity of an account increases (Figure~\ref{fig:partial-dependence-user-toxcity}) before decreasing again on the right side of the political spectrum. We thus find that when users mention users on the political extreme, this does not indicate increased toxicity; rather we find in general that users who reference these users tend to tweet less toxic content on Twitter. This may do with the tendency that the users who reference these politically polarized/extreme users also tend to be near the political extremes themselves. Creating a mention/@  graph among our 43,151~users, we find a moderate degree of assortativity (0.266), thus finding that users, on the whole, tend to interact with other users of similar political views (Figure~\ref{fig:graph-of-political-interactions}) and that this tendency is not necessarily correlated with increased toxicity.  Graphing the average partisanship of a user's mention against their own partisanship we further observe a high assortativity (Pearson correlation of $\rho=0.605$). 
\begin{figure}
\begin{minipage}[l]{0.48\textwidth}
\includegraphics[width=1\columnwidth]{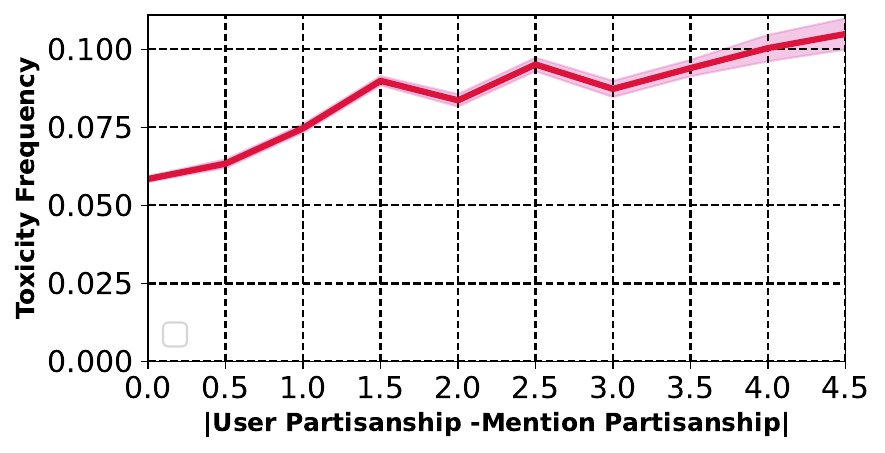}
\end{minipage}
\begin{minipage}[l]{0.33\textwidth}
\caption{As the difference in the partisanship of users and those that they mention/@ increases, the probability of users tweeting toxicly increases. 95\% Normal Confidence Intervals.
\label{fig:toxicity_vs_mention-new}}
\end{minipage}
\end{figure}

Instead, as was seen in (Figure~\ref{fig:partial-dependence-user-toxcity}) it is the difference in partisanship between a user and their mentions that linearly determines the toxicity of users. The average difference in the partisanship between a user and their mentioned accounts has a $\rho = 0.287$ Pearson correlation with the user's own toxicity and has a permutation importance of 0.080. Indeed, as seen in Figure~\ref{fig:toxicity_vs_mention-new}, we observe across our entire dataset that as the difference between a user's partisanship and the partisanship of the corresponding user that mention/@ increases the probability that they tweet toxicly increases. This illustrates, as found elsewhere~\cite{hanley2023sub,mamakos2023social}, that as users interact with more users different in partisanship than themselves, they are more likely to be toxic.  As an example a left-wing user (-1.53) with a particularly high standard deviation for the diversity of their mentions (1.818), often engaging in heated discussion with right-wing and left-wing accounts wrote the tweet concerning the former Republican US president Donald Trump:
\begin{displayquote}
\small
\textit{
This is so indescribably fucked up. Except I love Nancy Pelosi giving him the shiv.}
\end{displayquote}
\noindent
Similarly, a different  left-wing account (-1.504), which also regularly interacts with right-wing and left-wing accounts (1.65), regarding former Republican US president Donald Trump's son wrote:
\begin{displayquote}
\small
\textit{
Fuck him. No, seriously, fuck him. If anyone’s a welfare queen it’s him...}
\end{displayquote}
\noindent

\begin{figure}
\begin{minipage}[l]{0.48\textwidth}
\includegraphics[width=1\columnwidth]{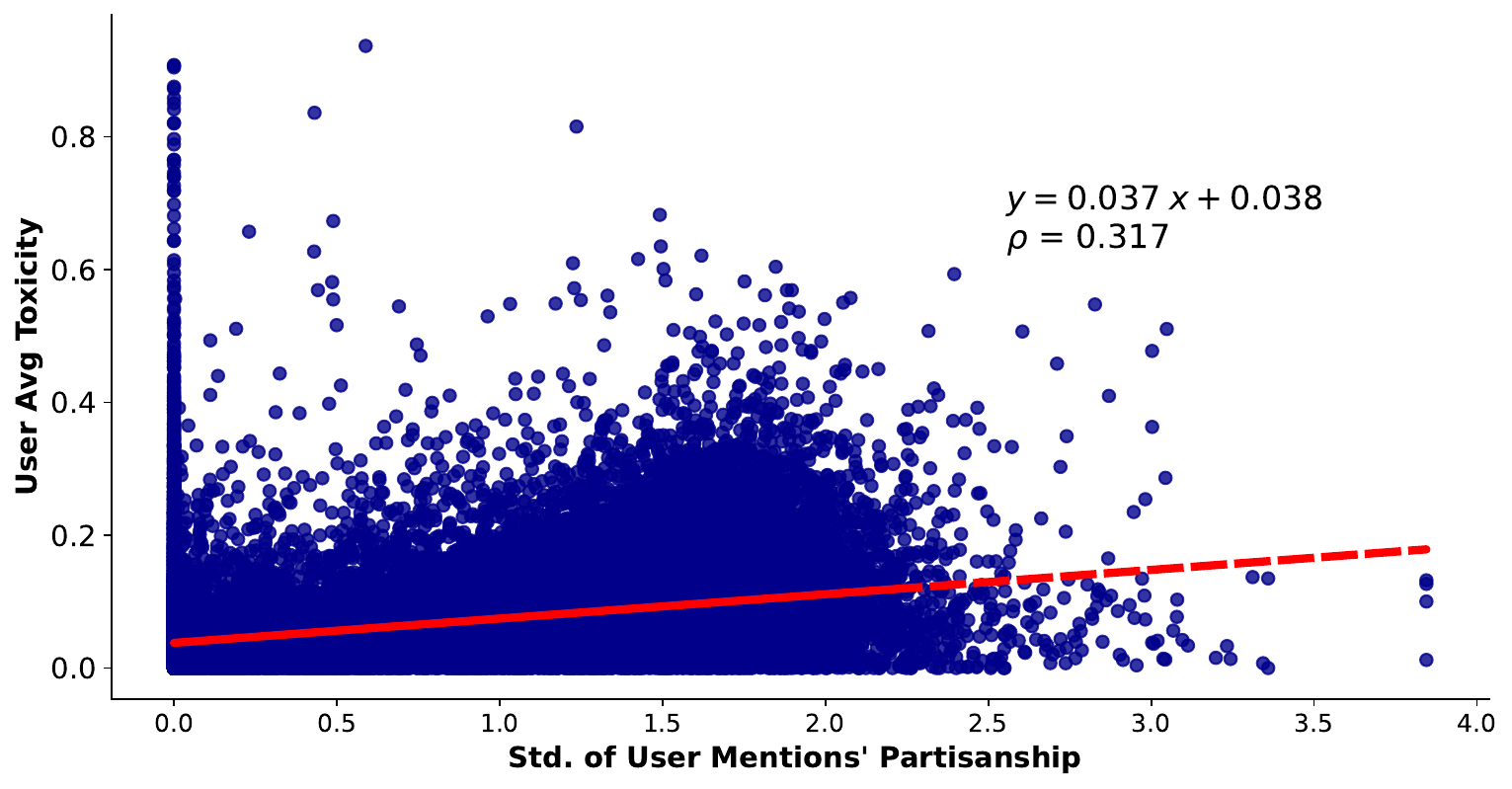}
\end{minipage}
\begin{minipage}[l]{0.33\textwidth}
\caption{As users mention a wider range of users along the political spectrum they are more likely to tweet toxic messages. 
\label{fig:politican-variance-toxicity2}}
\end{minipage}
\end{figure}

\vspace{2pt}\noindent
\noindent
\textbf{The Political Diversity of Mentions.}
In addition to finding that as users interact with more users different than themselves, from Figure~\ref{fig:partial-dependence-user-toxcity} and Table~\ref{table:importance-user-toxicity}, we find that as users mention/@ a wider political diversity of users, the more toxic their own tweets. With a Pearson correlation of $\rho =0.317$ and a permutation importance of  0.15, we see that this feature is relatively important in our fit model with it heavily contributing to the prediction of a user's toxicity (Figure~\ref{fig:politican-variance-toxicity2}).  This reinforces the finding of Mamakos et~al.~\cite{mamakos2023social}. who also found that when users engage with both left-leaning and right-leaning accounts on Reddit, they are more likely to engage in toxic behaviors on the platform.


\subsection{Summary}
In this section, using a GAM, we explored the role that several user-level characteristics have on the rate of user toxicity on Twitter. We find, most importantly, that users who interact and mention other users who regularly post toxic content are more likely to be toxic themselves. Similarly, we found the more a given user interacts with a politically diverse set of accounts, the more likely that account is to tweet toxic content. We replicate these results with the Perspective API in Appendix~\ref{sec:perspective-user-app} getting similar results. 

\section{Factors and Changes in Polarized and Toxic Topics on Twitter}
Having investigated the role that various user characteristics have in user toxicity on Twitter, we now explore how different characteristics affect different negative and toxic topics on Twitter. Specifically, how does the toxicity of topics on Twitter change based on the makeup of the user participating in these conversations? First discussing and performing some qualitative analysis on the most toxic and political ideological conversations on Twitter, we then determine how the political views, the diversity of political views, and the overall toxicity of the users participating in given conversations affected particular topics discussed in 2022.

\subsection{Setup}
In this section, we utilize a combination of MPNet and DP-Means as specified in Section~\ref{sec:topic-background} to perform topic analysis on the English language tweets within our dataset. After running our algorithm on the 5.5M~toxic tweets from our set of 43.1K~Twitter users, we identified 5,288~clusters with at least 50~toxic tweets. Upon identifying these clusters, as outlined in Section~\ref{sec:topic-background}, we further extract the most characteristic (often offensive) words within each cluster as well as each cluster's most representative toxic tweet. Before further detailing some of the characteristics of each of these toxic tweet clusters, we now give a brief overview of how we estimate the overall toxicity and political bend of each particular topic after identifying their corresponding cluster of toxic tweets.

\begin{figure}
\begin{subfigure}[l]{0.45\textwidth}
\includegraphics[width=1\columnwidth]{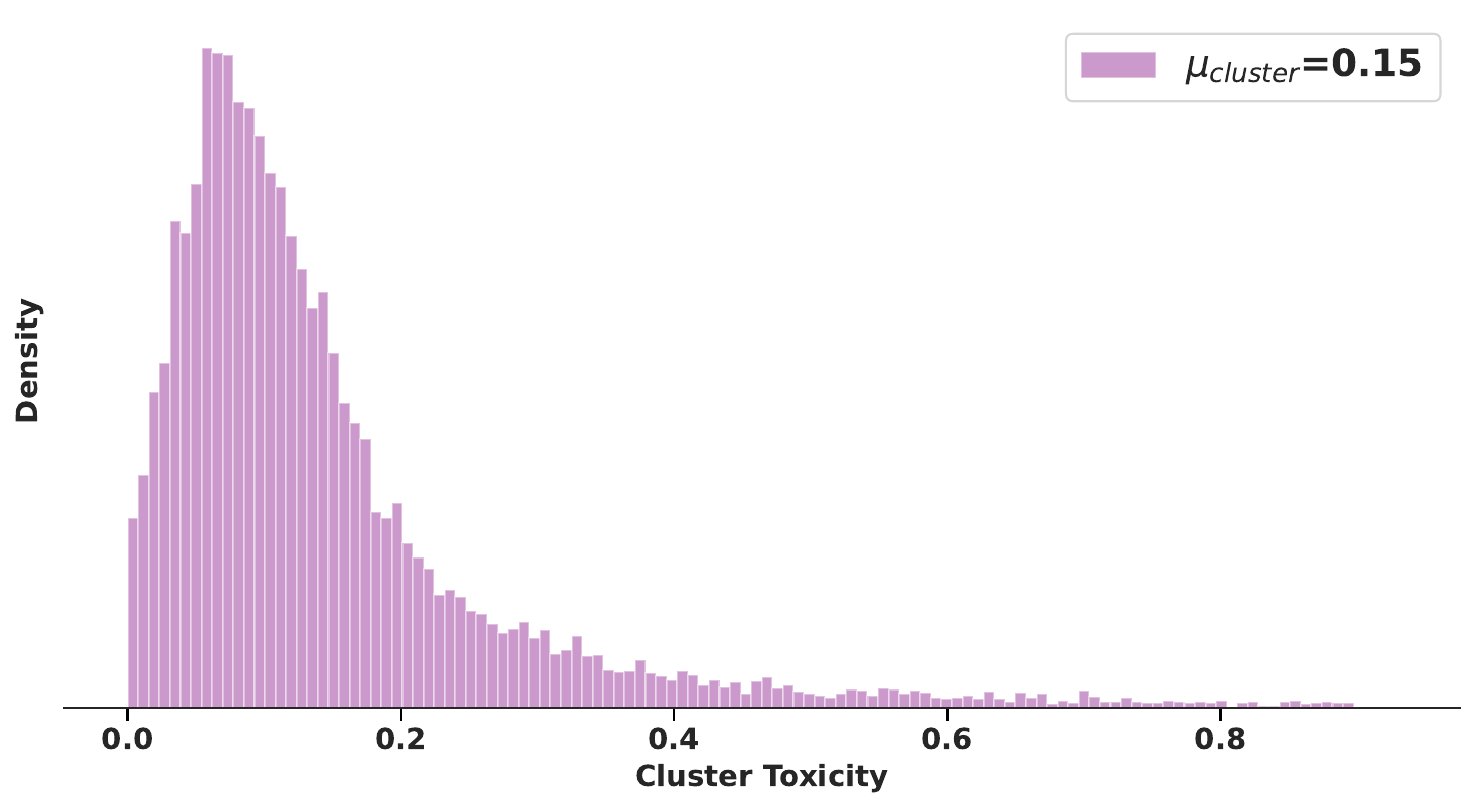}
\caption{}
\label{fig:cluster-dists-toxic}
\end{subfigure}
\begin{subfigure}[l]{0.45\textwidth}
\includegraphics[width=1\columnwidth]{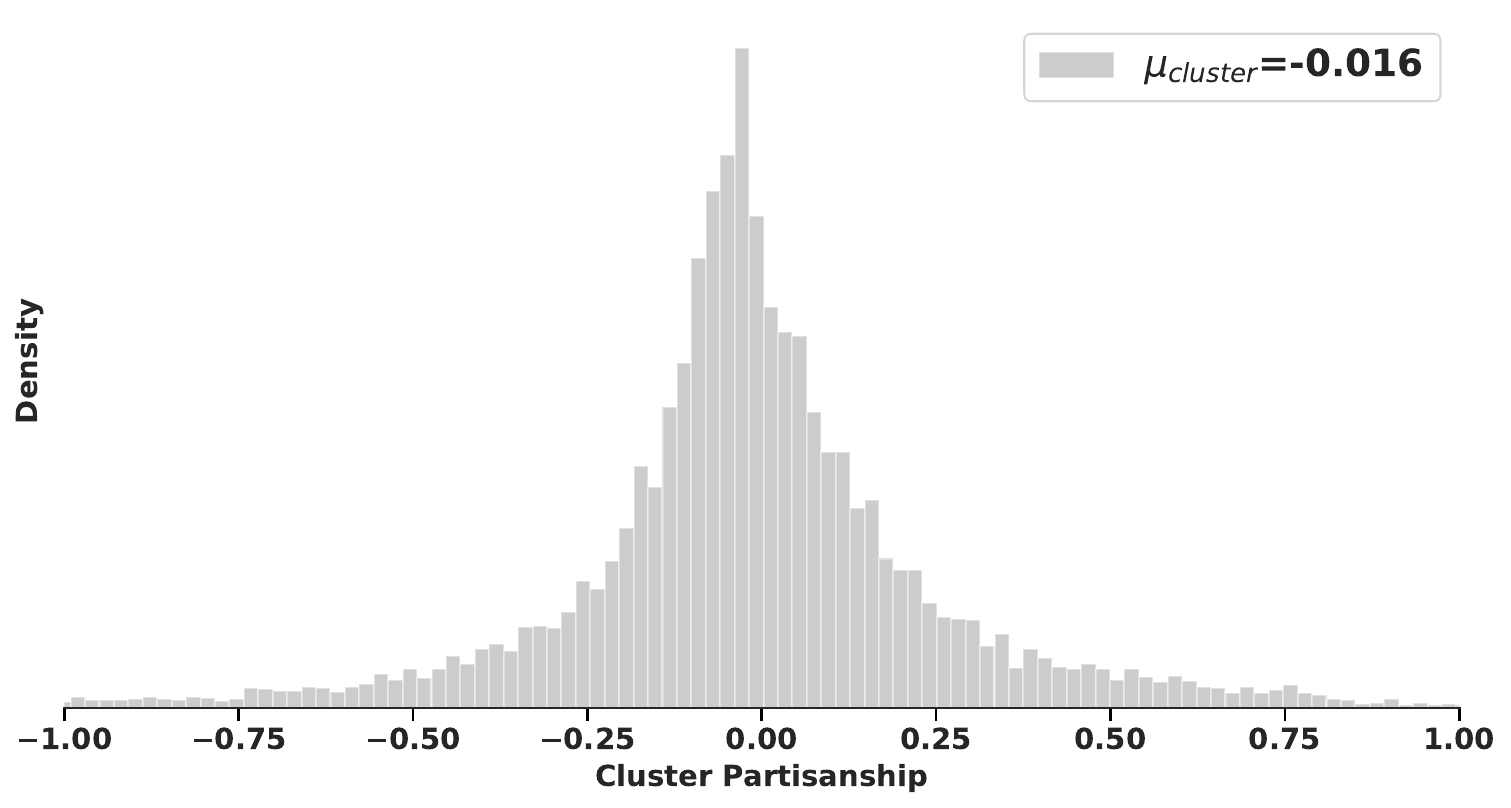} 
\caption{}
\label{fig:cluster-dists-part}
\end{subfigure}
\caption{The distribution of toxicity and partisanship within our set of clusters.}
\end{figure}

\vspace{2pt}\noindent
\noindent
\textbf{Estimating the Toxicity of Topics.}
To estimate the toxicity of particular topics, we determine the average toxicity score of all tweets present within that given cluster. While we largely rely on our average toxicity scores, in addition to this metric, we further determine the \emph{percentage} of toxic tweets within our \emph{entire} English-language dataset that conforms to that particular topic.  Namely, after identifying each toxic cluster center, for each of these toxic cluster centers, we further identify the set of non-toxic tweets that also conform to the topic. We then calculate the percentage of toxic tweets (\textit{i.e.}, toxicity > 0.5) per topic. 

To assign non-toxic tweets to our set of toxic tweet centers, we utilize the approach laid out in prior work~\cite{hanley2022happenstance,hanley2023partial} and subsequently assign each non-toxic tweet to the cluster center with the highest semantic similarity to the tweet. As recommended by Hanley {et~al.}~\cite{hhanleyspecious2024}, given our fine-tuned version of MPNet, we again utilize a cluster threshold of 0.60 for assigning a given non-toxic tweet to a given cluster. We plot the distribution of estimated topic toxicity in Figure~\ref{fig:cluster-dists-toxic}. We utilize this approach, rather than clustering all 89.6 million English tweets given the size of our dataset, and because, for this work, we largely are only concerned with topics that have some level of toxicity. 

\vspace{2pt}\noindent
\noindent
\textbf{Estimating the Partisanship of Topics.}
To further examine the role of partisanship within interactions within particular topic clusters, we further determine the overall political orientation of each cluster. To do so, after assigning all remaining non-toxic tweets to our clusters as specified above, we subsequently determine which set of users participated/tweeted about that topic. Calculating the average and standard deviation of the political orientations of all the Twitter users (utilizing our previous calculations of user partisanship [Section~\ref{sec:background-ca}]) that tweeted about that topic, we thus estimate each topic's political-ideological composition. We plot the distribution of the partisanship of our set of clusters in Figure~\ref{fig:cluster-dists-part}.

\subsection{The Most Toxic Topics of 2022\label{sec:most-toxic}}

\begin{table*}
\centering
\scriptsize
\selectfont
\setlength{\tabcolsep}{4pt}
\begin{tabularx}{\textwidth}{lXrrrXrrr}
\toprule
 &   &  &   \# Toxic & Avg.& Example & Avg.  & Avg. Partisan  & Partisan \\
 Topic& {Keywords}&\# Tweets &Tweets  & Toxicity &  Tweet  & Partisan. & of Toxic Users& Std. \\
\midrule
 1 & biden, joe, administration, president, senile  &246,868 & 39,102 (15.84\%) & 0.1824& Joe Biden And everything is screwed up. You suk & 0.642 &0.379 &1.084\\ 

 2 & ukraine, russia, kyiv, putin, independent &763,153 & 36,425 (4.77\%) &0.070 &  So l guess you what Ukraine to stop fighting back and let the Russians kill them. Ukraine Will Resist Fuck Putin& -0.091 &0.031 & 0.899\\ 

3 & lie, pathological, truth, habitual, liar  &100,825 & 26,894 (26.67\% ) &0.298 & These leftist serial liars always project onto others the crimes they are perpetrating. & -0.055 & 0.081 & 1.0300  \\

 4 & party, democrat, republican, dnc, destroying &111,763 & 22,705 (20.32\%) &0.232 & That slate is FAR better than the gaggle of corrupt Marxists the racist lunatic democrat party pushed forward. Nobody is gonna give you a nod for badmouthing the better team.
  &0.215&  0.171 & 1.162\\

 5& ballot, election, stolen, voting, rigged& 295,356 &  22,399 (7.58\%) &0.093 & You already know that the Maricopa County Election will say "Fuck Your Ballots" and ram it through the certifications.& 0.199 & 0.121 & 1.131 \\

  6 & fox, news, murdoch, carlson, tucker & 179,224 & 20,958 (11.69\%) &0.138 &Fox News Give it a rest already. For is even worse than national enquirer for false made up trash. & -0.027 & 0.089 & 1.116 \\

  7 & filipkowski, ron, flynn, bannon, nut & 107,255& 19,018 (17.73\%)&0.187 & @REDACTED Man are Fox ppl nuts or what &-0.686 & -0.535 & 0.784  \\

  8 & tweet, follow, deleted, algorithm, account & 190,665 & 18,822 (9.87\%)  &0.102& Ok is anybody else's twitter completely fucked up and glitchy?
 & 0.044 &  0.038 & 0.982\\

  9 & stupidity, smart, intelligent, educated, dumb & 23,232 & 18,533 (79.77\%)&0.688 &@I mean how stupid are these people or what?
What happened to like history classes?
Gees what a bunch of loser white people.&  0.141 &  0.131 & 0.938 \\

  10 & abortion, birth, pro-life, pregnancy, fetus &309,723 & 18,456 (5.96\%)&0.087& This moron thinks the Supreme Court literally edited the Bill of Rights to remove the right to an abortion. Dumb as fucking rocks these people.
 &  0.114 &.112 & 1.265 \\

\bottomrule
\end{tabularx}
\caption{Top toxic topics---by the number of toxic tweets---in our dataset.\label{table:toxic-topics}} 
\end{table*}
We start this section by providing an overview of the topics with the most toxic tweets in 2022 (Table~\ref{table:toxic-topics}). We further give an overview of the most partisan topics in Appendix~\ref{sec:partisan-toxic} and the most toxic topics in Appendix~\ref{sec:most-toxic-by-percentage} (most of these topics are merely users calling each other different epithets). As seen in Table~\ref{table:toxic-topics}, many of the most common toxic tweets concerned the most politically divisive issues of 2022~\cite{Montanaro2022}, namely, Joe Biden's administration (Topic 1;~246,968K tweets), Russia's invasion of Ukraine (Topic 2;~763K tweets), and the abortion rights in the United States in the wake of the Dobss v. Jackson decision which overturned US federal abortion rights~\cite{StaffAborrtion2022}.

Examining the average partisanship of the user who tweeted about each of the top toxic topics, we find distinct political differences. Markedly, we observe, that those who tweeted in a toxic manner about the Ukraine War tended to have a slight rightward tilt (+0.122 rightward tilt). Examining these tweets, we find right-leaning users when tweeting about the war, excoriated or derided the Ukrainian government or military, which was picked up as toxic by our contrastive-DeBERTa model. 
For example, one ``toxic'' tweet by a rightward user stated: 
\begin{displayquote}
\small
\textit{
No more arms for a Ukraine refusing to negotiate! Ukraine doesn't need more arms, Ukraine needs more intelligence! And Zelensky is a dictatorial asshole!}

\end{displayquote}
In contrast, considering all users who tweeted about the war, we find that they tended to lean leftward (-0.091 leftward tilt) with one left-leaning user tweeting:
\begin{displayquote}
\small
\textit{
Stand With Ukraine!}
\end{displayquote}
Looking at the users who tweeted about Joe Biden's presidency (Topic~1),  we again see a rightward bias (+0.642) among users who tweeted about him or his administration generally and with users who tweeted about him in a toxic manner (+0.379). For example, one user tweeted
\begin{displayquote}
\small
\textit{
Save the poor water bottle from that pedophile Joe Biden before he becomes a victim}
\end{displayquote} We thus observe that those talking about the administration (both in a toxic and non-toxic manner) were largely right-leaning (as largely expected given that the Biden administration is Democratic). Finally, examining the set of users who tweeted about abortion in 2022 (Topic~10), we again find a rightward lean among users who tweeted about this issue. For example, one right-wing user wrote:
\begin{displayquote}
\small
\textit{Why are actors so ignorant about policies? States can still do abortions. Go ahead and murder more babies.}
\end{displayquote}

Besides these politically salient issues, we observe several topics where politically charged users simply derided each other (Topic~4), called each other idiotic (Topic~9), or called the other political side liars (Topic~3). We further see in Topic~5 heavy emphasis on the US presidential election being stolen in Arizona. As documented by Prochaska et~al., a misinformation story called Sharpiegate where ``Sharpies invalidated ballots in Maricopa County, Arizona'' was widely spread on Twitter and we see evidence of it in our dataset with several political users heatedly and toxically calling the Arizona election rigged~\cite{prochaska2023mobilizing}.

\subsection{Topic Dependent Changes in Partisanship and Toxicity \label{sec:changes}} Having explored some of the most prominent toxic topics during our period of study, we now explore how the toxicity of different Twitter topics change as users of different political orientations enter and leave. We find that regardless of whether a topic moderates (\textit{i.e.}, political orientation moves closer to 0) or becomes more extreme (\textit{i.e.}, political orientation becomes more left-leaning or more right-leaning), on average, this movement has little bearing on toxicity. Indeed correlating the change in the political orientation of a given topic between January and December with the percentage change in the toxicity of that conversation, we calculate a Pearson correlation of $\rho=-0.0168$, indicating little to no relationship. Similarly, we find that the variance of political participation in particular topics over time is also only slightly correlated with the toxicity of a given topic $\rho=-0.098$. This indicates that unlike for users, a different dynamic may be influencing the toxicity of particular topics across time. 

Across our dataset, we find that regardless of whether the topic moderates or moves to the extremes, in both cases, toxicity generally increases (55.8\% of the time for topics that moderated in partisanship and 71.4\% of the time for topics that moved to the political extreme). Furthermore, we find that between January 2022 and December 2022, in 34.8\% of topics, as topics became more right-leaning, they also became more toxic; in 27.1\% cases, they became less toxic as they became more right-leaning. Conversely, in 21.2\% of our topics, they became more toxic as they became more left-leaning, and in  17.0\% of topics they became less toxic as they became more left-leaning. However, examining each cluster, we \emph{do} find that on a cluster-by-cluster basis as the political composition of users involved in that topic changes there are corresponding changes in toxicity.

\begin{figure}
\begin{subfigure}[l]{0.32\textwidth}
\includegraphics[width=1\columnwidth]{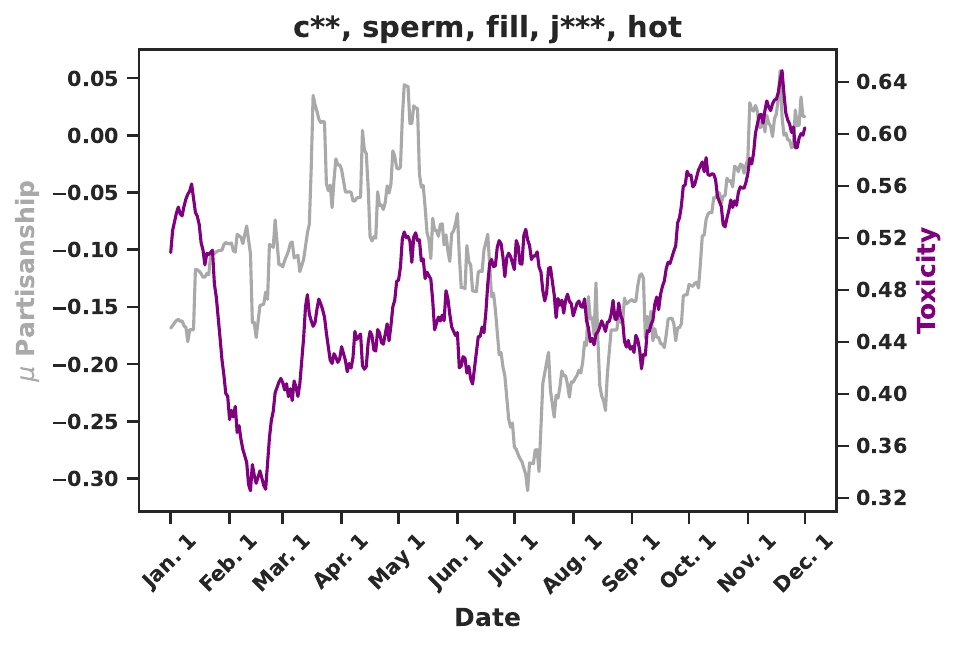} 
\caption{}
\label{fig:cum}
\end{subfigure}
\begin{subfigure}[l]{0.32\textwidth}
\includegraphics[width=1\columnwidth]{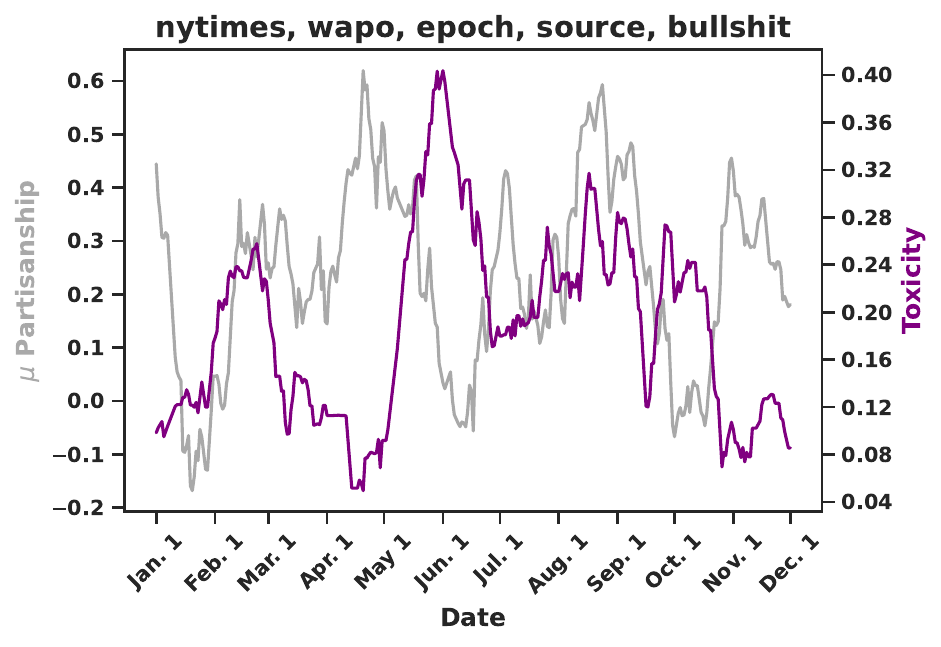} 
\caption{}
\label{fig:nytimes}

\end{subfigure}
\begin{subfigure}[l]{0.32\textwidth}
\includegraphics[width=1\columnwidth]{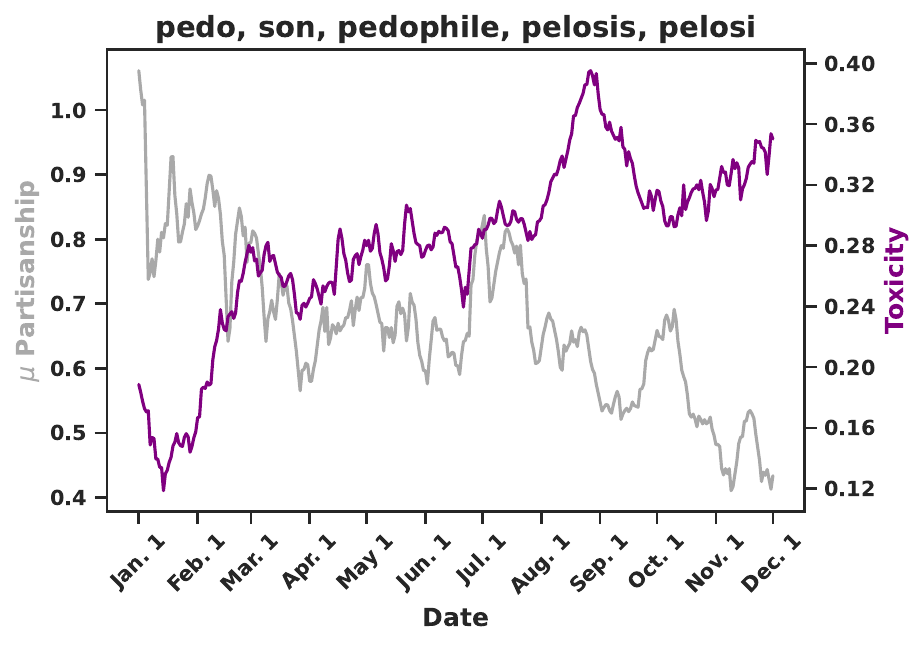} 
\caption{}
\label{fig:pedo}
\end{subfigure}
\begin{subfigure}[l]{0.32\textwidth}
\includegraphics[width=1\columnwidth]{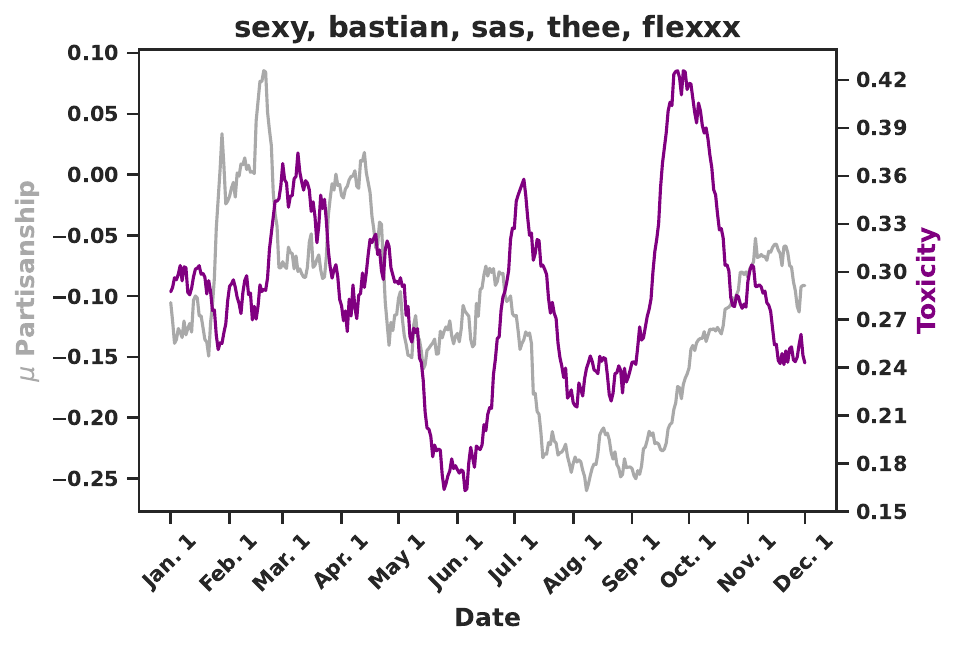} 
\caption{}
\label{fig:sexy}
\end{subfigure}
\begin{subfigure}[l]{0.32\textwidth}
\includegraphics[width=1\columnwidth]{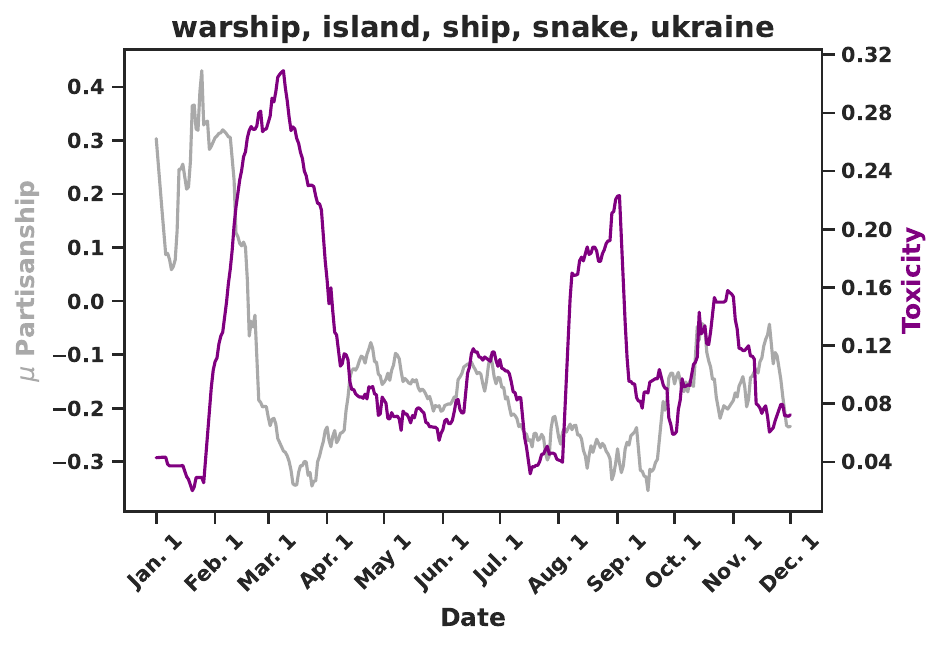} 
\caption{}
\label{fig:warship}
\end{subfigure}
\begin{minipage}[l]{1\textwidth}
\caption{Topics with the largest increase in toxicity in 2022. \label{fig:toxicity-swing}}
\end{minipage}
\vspace{-15pt}
\end{figure}

\begin{figure}
\begin{subfigure}[l]{0.32\textwidth}
\includegraphics[width=1\columnwidth]{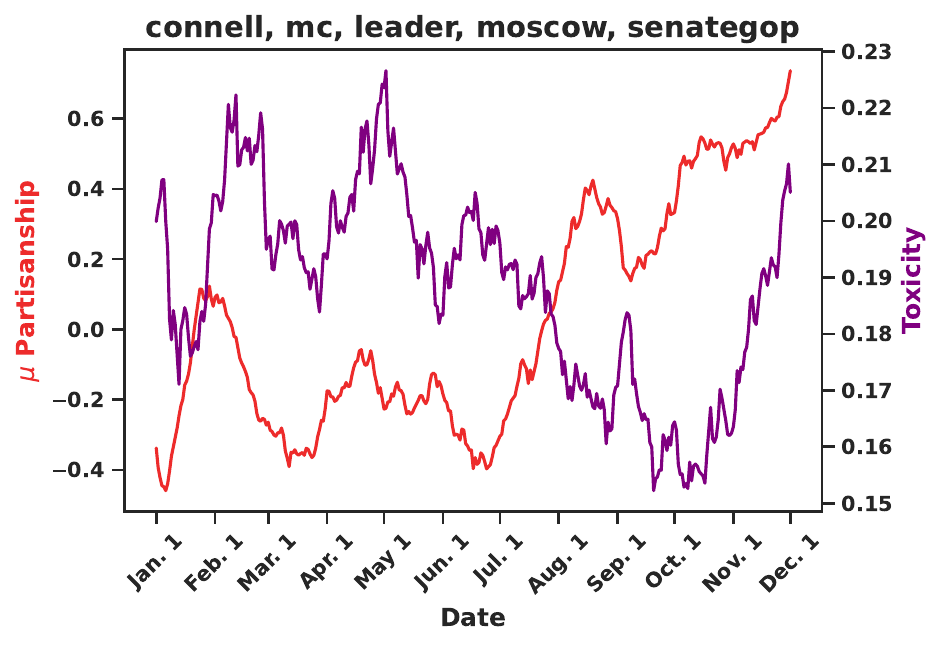} 
\caption{}
\label{fig:moscow}
\end{subfigure}
\begin{subfigure}[l]{0.32\textwidth}
\includegraphics[width=1\columnwidth]{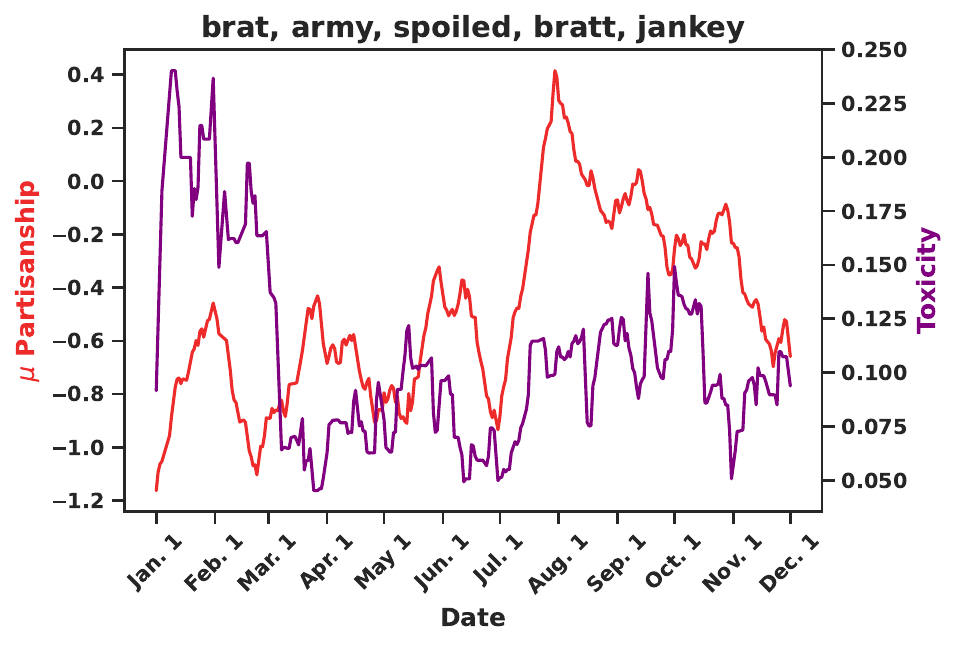} 
\caption{}
\label{fig:brat}
\end{subfigure}
\begin{subfigure}[l]{0.32\textwidth}
\includegraphics[width=1\columnwidth]{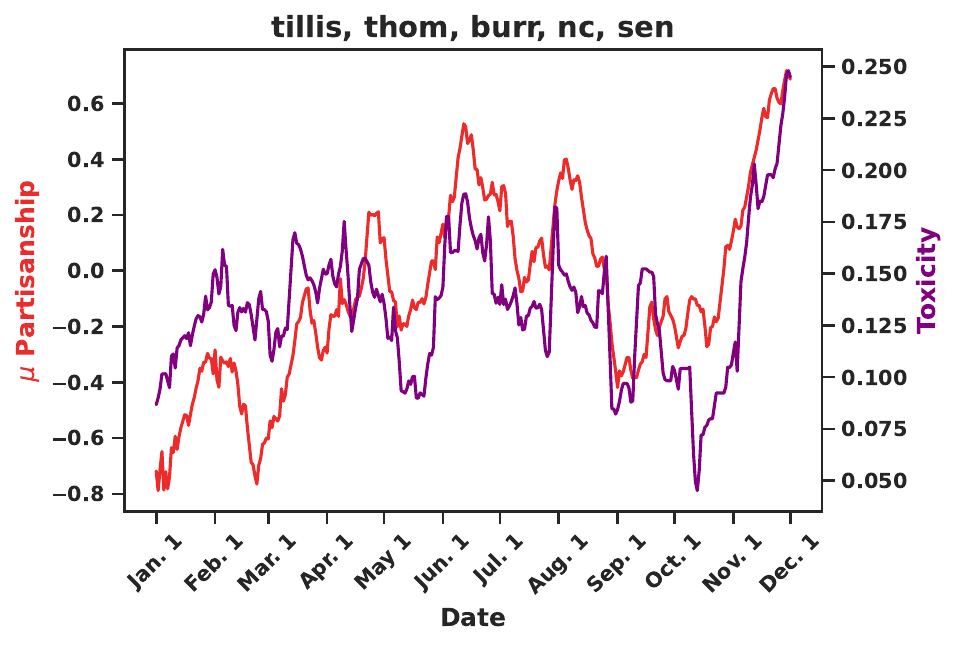} 
\caption{}
\label{fig:tillis}
\end{subfigure}
\begin{subfigure}[l]{0.32\textwidth}
\includegraphics[width=1\columnwidth]{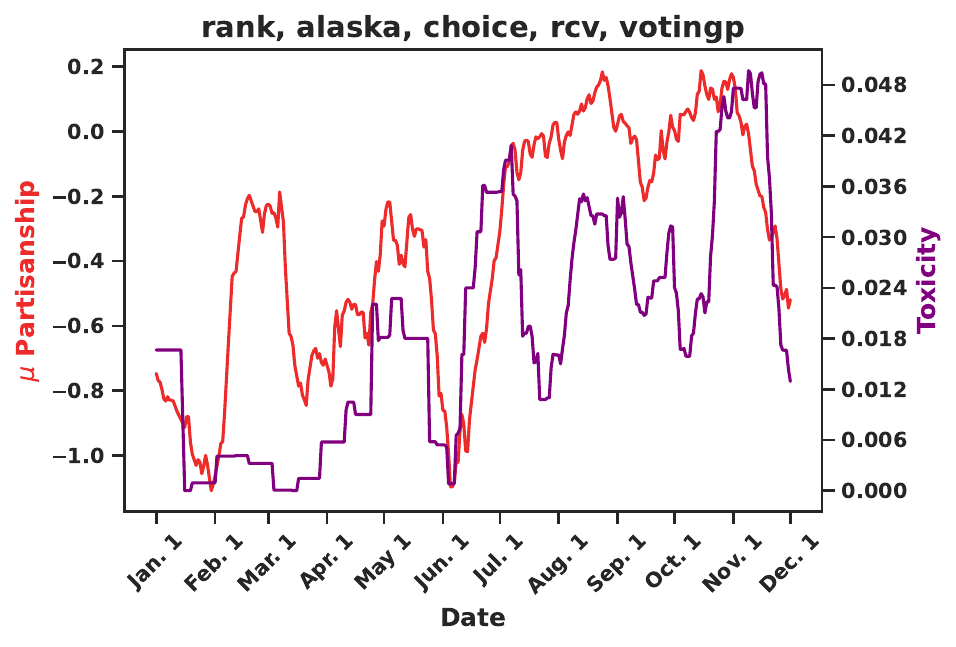} 
\caption{}
\label{fig:rank}
\end{subfigure}
\begin{subfigure}[l]{0.32\textwidth}
\includegraphics[width=1\columnwidth]{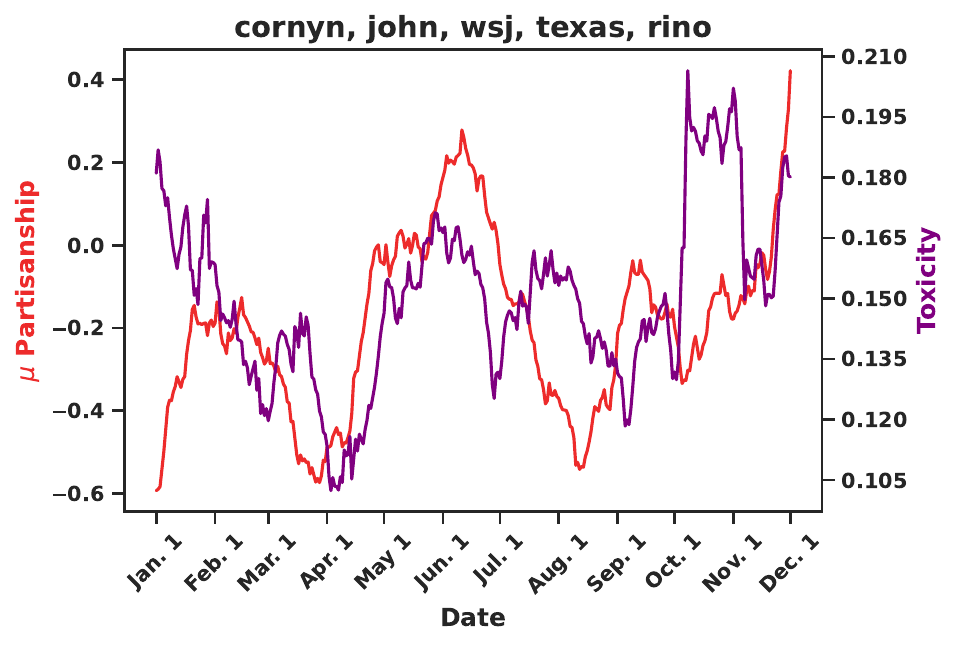} 
\caption{}
\label{fig:cornyn}
\end{subfigure}
\begin{minipage}[l]{1\textwidth}
\caption{Topics with the largest swing to right-leaning partisanship throughout 2022. \label{fig:conservative-swing}}
\end{minipage}
\end{figure}

\vspace{2pt}
\noindent
\textbf{Toxic Swings.} To further qualitatively understand the nature of how toxicity and political orientation change over time, we plot the toxicity and partisanship for the topics with the largest increases in toxicity between January 2022 and December 2022. We observe that while for four topics considered, (Figures~\ref{fig:cum},~\ref{fig:nytimes},~\ref{fig:sexy}, and~\ref{fig:warship}) as the topic became more right-leaning, toxicity similarly increased, for one of the topics (Figure~\ref{fig:pedo}), we observe the opposite. Examining, each we observe, noticeable trends where, depending on the political nature of the topic,  a corresponding swing in the political composition of the users in the left or the right direction, is correlated with an increase in toxicity. For instance, in the tweets surrounding the New York Times and the Washington Post's accuracy,  we observe that as users discussing the topic became more right-leaning, the more toxic the tweets. We similarly find for the topic surrounding the destruction of Russian warship on Snake Island by Ukrainians, the more right-wing the users, the more toxic. For example, one user wrote:
\begin{displayquote}
\small
\textit{
    Surprising Russian Navy Losses Against Ukraine Century After Tsushima 
Ukraine is really FUCKING Russian Navy Ship's up during the Russian Invasion into Ukraine}
\end{displayquote}

\noindent
In contrast, for Topic 3 (Figure~\ref{fig:pedo}), we observe that as users became more left-leaning the overall toxicity of the topic decreased. We observe that this is largely due to left-leaning users adopting retorts to right-leaning users calling the Democratic former Speaker of the House Nancy Pelosi a pedophile. For example, we observe one user stating:

\begin{displayquote}
\small
\textit{
Let's not forget that the last republican speaker of Michigan house was a Pedophile who raped a 15 year old sister in law.}
\end{displayquote}

\noindent
We thus observe among these top topics that depending on the political nature of the given topic and how users are interacting and replying to other users about the topic, a corresponding swing in the political composition of the users in the opposite direction, may or may be correlated with an increase in toxicity.

\begin{figure}
\begin{subfigure}[l]{0.32\textwidth}
\includegraphics[width=1\columnwidth]{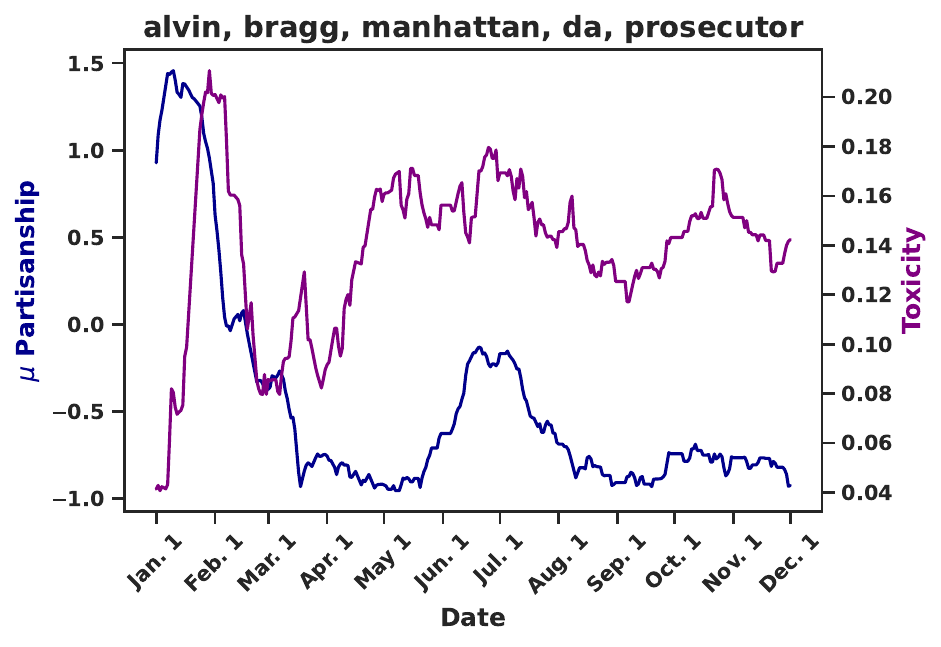} 
\caption{}
\label{fig:alvin}
\end{subfigure}
\begin{subfigure}[l]{0.32\textwidth}
\includegraphics[width=1\columnwidth]{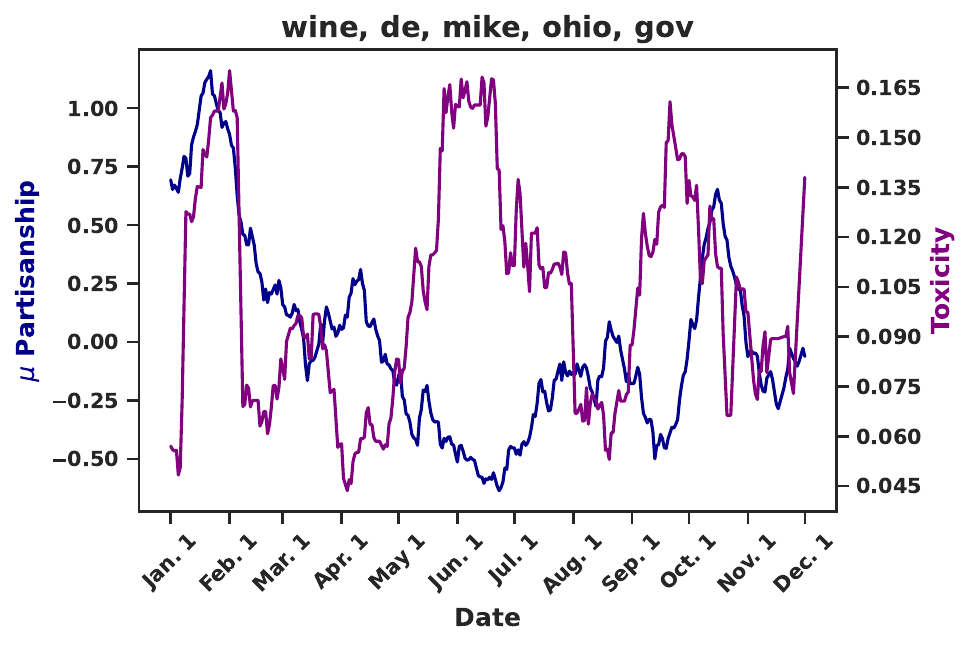} 
\caption{}
\label{fig:wind}
\end{subfigure}
\begin{subfigure}[l]{0.32\textwidth}
\includegraphics[width=1\columnwidth]{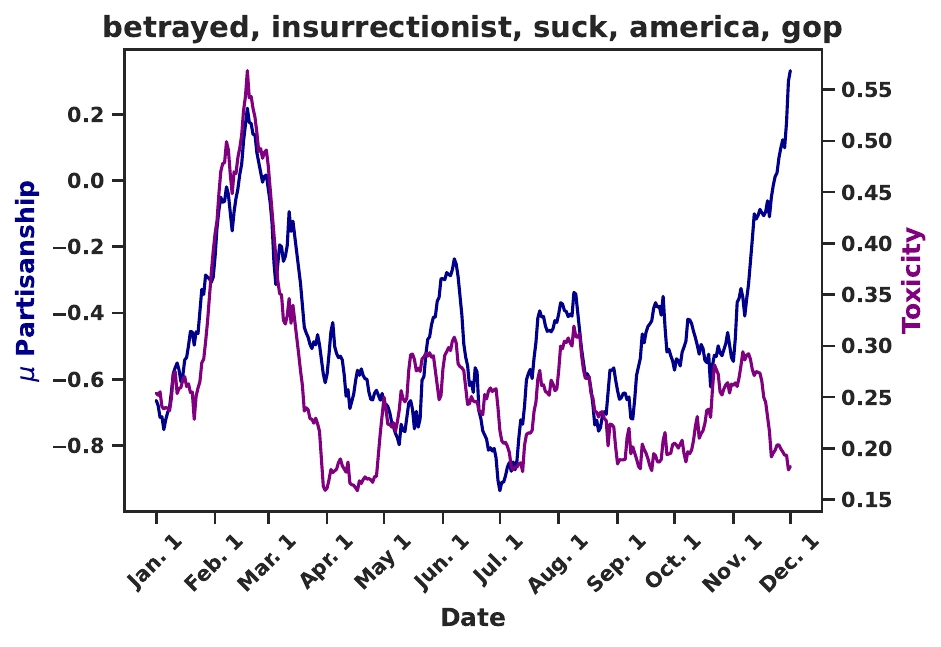} 
\caption{}
\label{fig:betrayed}

\end{subfigure}
\begin{subfigure}[l]{0.32\textwidth}
\includegraphics[width=1\columnwidth]{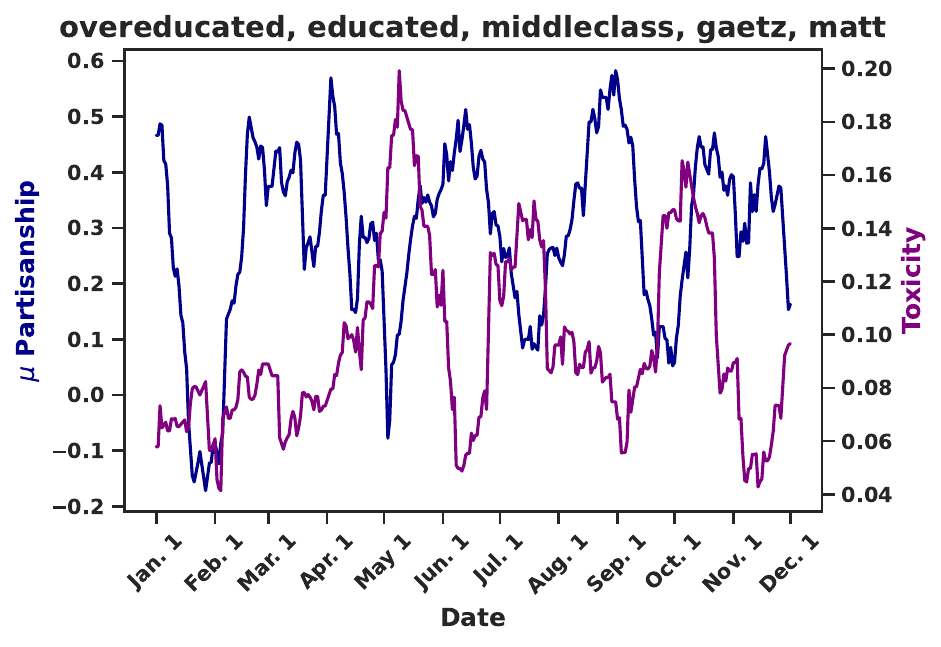}
\caption{}
\label{fig:overeducated}
\end{subfigure}
\begin{subfigure}[l]{0.32\textwidth}
\includegraphics[width=1\columnwidth]{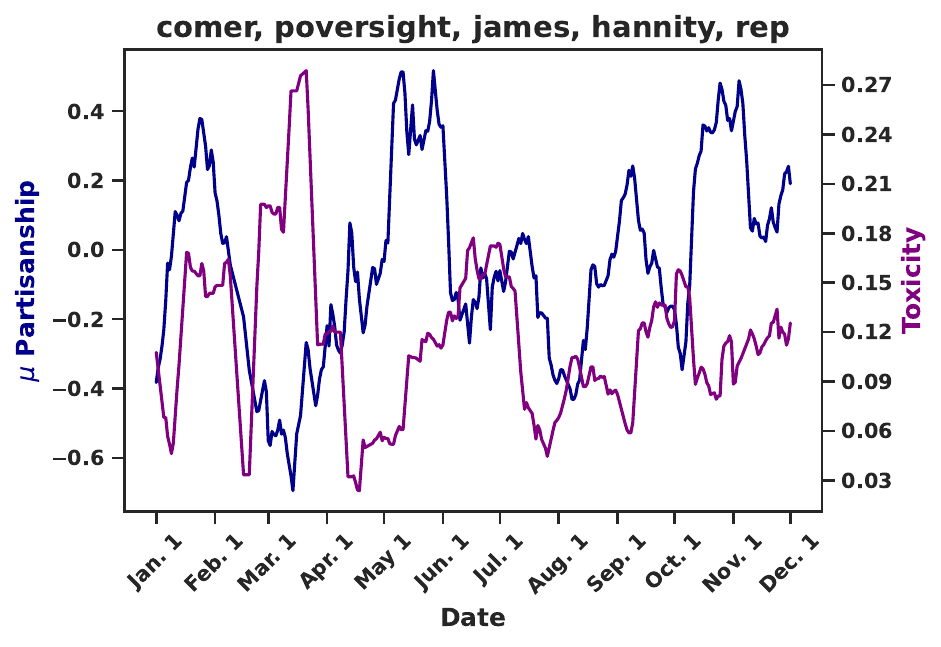} 
\caption{}
\label{fig:comer}
\end{subfigure}

\begin{minipage}[l]{1\textwidth}
\caption{Topics with the largest swing to left-leaning partisanship throughout 2022. \label{fig:liberal-swing}}
\end{minipage}

\end{figure}

\noindent
\vspace{2pt}
\textbf{Left-Leaning and Right-Leaning Swings.} Plotting the set of topics with the largest swings in average political orientation, to both the right and left-leaning end, between January 2022 and December 2022 (Figures~\ref{fig:conservative-swing} and~\ref{fig:liberal-swing}), we again observe that changes in toxicity as a result of these changes are largely dependent on the topic. For example, as the conversation surrounding Tom Tills (the senior Republican Senator for North Carolina) became more right-leaning, the toxicity of that topic increased dramatically (Figure~\ref{fig:tillis}). Despite Senator Tillis being a Republican, we observe that this is largely due to right-leaning users largely labeling Senator Tillis a RINO (Republican in name only) with one user posting:

\begin{displayquote}
\small
\textit{You've always been a RINO
NC must be ashamed of you}
\end{displayquote}
\noindent We find a similar behavior for Senator John Cornyn of Texas, again with a user writing:

\begin{displayquote}
\small
\textit{John Cornyn This Bill is trash. RINOs need to go.
Cornyn votes with the Democrats almost as often as his own party.
Texas should be ashamed}
\end{displayquote} We similarly find that as right-leaning users joined the conversation about US Senate Republican Minority Leader Mitch McConnell being beholden to the Russian government~\cite{Hulse2018} toxicity increased. We note that the attacks against Senators Mitch McConnell, John Cornyn, and Tom Tillis were \emph{all} largely for not being conservative enough. Finally, for Democratic Manhattan District Attorney Alvin Bragg, we also observe that when more right-leaning users joined the conversation surrounding him, toxicity increased. However, unlike for the Republican Senators, more intuitively, this was largely due to his investigation of former Republican President Donald Trump. 

In contrast, for Republican Ohio Governor Mike DeWine, we observe that as more left-leaning users joined the conversation surrounding him the topic became more toxic, with one user writing
\begin{displayquote}
\small
\textit{
Gov Mike DeWine Thank you, Gov Mike DeWine, for making it easier for Ohioans to be killed by gun violence. Fuck you.}
\end{displayquote}
Similarly, for Republican Florida Congressman Matt Gaetz, we also observe that as more liberal users joined the discussions surrounding him the topic became more toxic. We find that this was largely sparked by a tweet from Matt Gaetz stating:

\begin{displayquote}
\small
\textit{Over-educated, under-loved millennials who sadly return from protests to a lonely microwave dinner with their cats, and no bumble matches.}
\end{displayquote}
\noindent to which one user replied 
\begin{displayquote}
\small
\textit{Only stupid, insecure men worry about women being over-educated. Which one are you, matt gaetz?}
\end{displayquote}

\noindent
We thus observe that the context of each of these topics, in particular, is decisive for determining how different swings in political polarization will affect the overall toxicity of the topic. As within individual users (See Section~\ref{sec:toxic_middle}), partisanship itself does not necessarily predict a higher degree of toxicity within conversations but is largely topic-dependent. Even the target/topic being a right-leaning or left-leaning entity/individual not decisively giving whether a left or right-leaning shift in users will correspond to an increase in toxicity. 

\subsection{Topic User Composition and the Toxicity of Topics\label{sec:topic-level-gam}} Having qualitatively described the composition and changing dynamics of some of our set of topic clusters, we now determine how the several user-level features of individual topic clusters predict the toxicity within the topic to better understand what may be influencing the toxicity of individual topics.

We note, and as seen throughout this section, topics on Twitter vary widely with individual topics often varying widely in political composition over time. Across all topics considered in our dataset, on average between January 2022 and December 2022, the political composition of the users tweeting about each topic changed by 0.159 standard deviations (based on the latent space that we previously determined [Section~\ref{sec:background-ca}]). In 61.9\% of cases, topics became more right-leaning, and in 38.1\%  topics became more left-leaning; similarly, within this same period, 56.0\% became more toxic while 44.0\% became less toxic. As a result, to quantify the effect that the composition of users has on the toxicity of a given topic at a single point in time, for each topic and each month combination, we gather the user compositions and the cluster characteristic data. We thus, in this section seek to determine the factors that determine the average toxicity score of a topic within a single month time-span. 

As before, to determine the role of various topic-level features in the overall toxicity of that cluster, we fit a GAM on the average toxicity score each month within each of our clusters against: \begin{enumerate}
    \item \emph{The number of users who tweeted about that topic.}
    \item \emph{The average user toxicity in the cluster.}
    \item \emph{The percentage of users involved in that topic that is Twitter verified.}
    \item \emph{The average of the partisanship in that cluster.}
    \item \emph{the standard deviation of political ideologies of users within that topic cluster.}
    \item \emph{The average age of the users in clusters.}
\end{enumerate}

Again, as in Section~\ref{sec:toxic_middle} when fitting our model, we perform variable selection based forward selection based on the Akaike Information Criterion~\cite{akaike2011akaike} Furthermore, again, to ensure that our model generalizes, we reserve out 10\% of our data as validation, and in our results report our model's $R^2$ value on this validation set. After fitting this regression, we further determine the estimated importance of each variable to our final model by permuting the features and seeing the estimated impact on the $R^2$ score of the validation set of our data. We do not consider other user account characteristics due to their multicollinearity with user toxicity (as seen in Section~\ref{sec:toxic_middle}, many user characteristics are correlated with their individual toxicity). Finally, we again reproduce our results with the Perspective Toxicity API in Appendix~\ref{sec:cluster-perspective-tox} obtaining similar results.

\begin{figure}
\begin{minipage}[l]{1.0\textwidth}
\includegraphics[width=1\columnwidth]{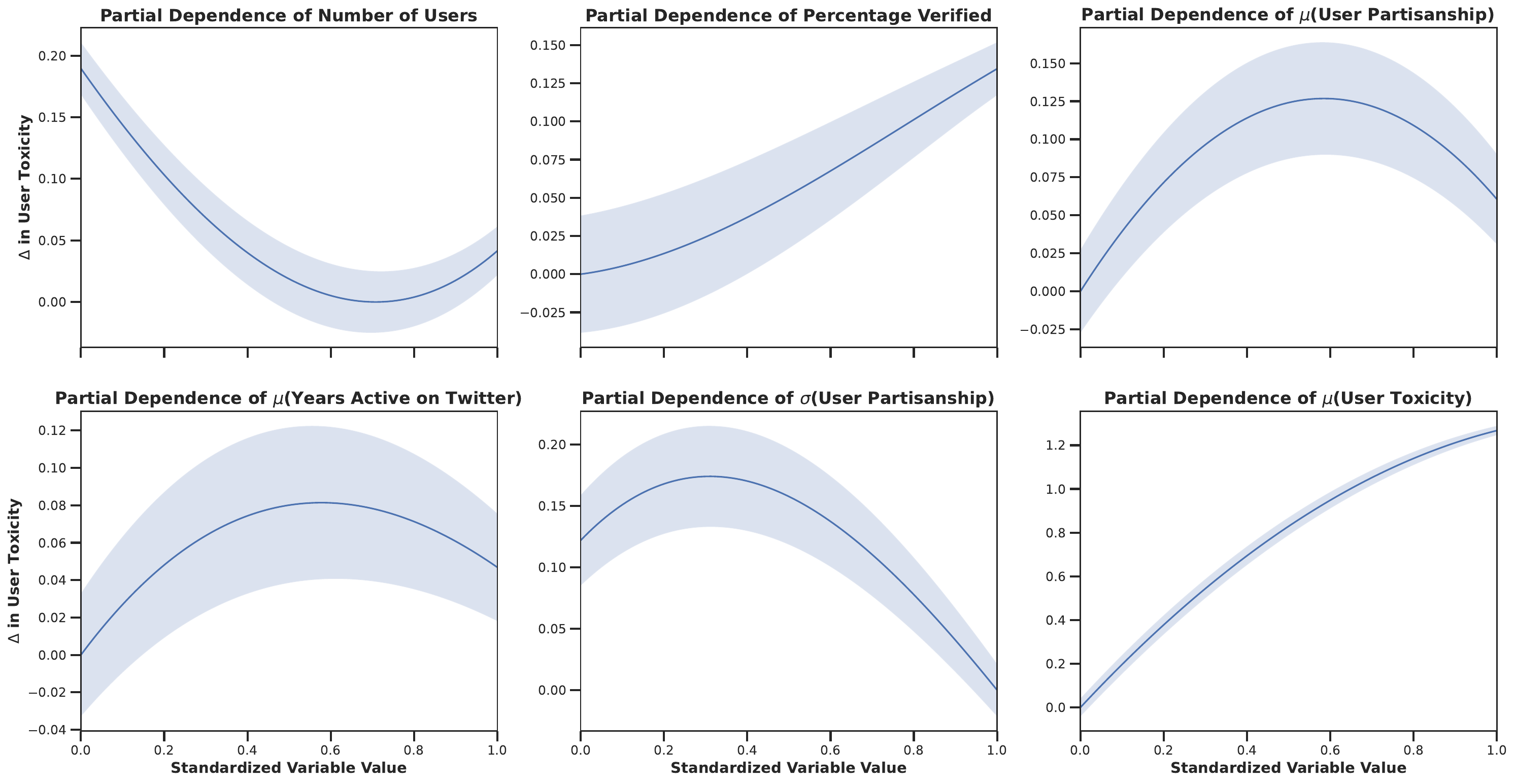} 
\end{minipage}
\begin{minipage}[l]{1\textwidth}
\caption{Partial dependencies with 95\% Normal confidence intervals between our fitted standardized dependent variables and cluster toxicity.}
\label{fig:partial-dependcies-clusterlevel}
\end{minipage}

\end{figure}

\begin{table}
    \small
    \centering
    \begin{tabularx}{0.85\columnwidth}{l|rrr}
     Train $R^2$: 0.397, Validation $R^2$:  0.389  \\
    \toprule
      Dependent Variable  & Pearson Corr. $\rho$ & Kendall's $\tau$ & Permut Import. \\    \midrule
  Number of Users &-0.292 & -0.139  & 0.445  \\
 $\mu$(Years Active on Twitter) &-0.191 & -0.186 & 0.018\\
  Percentage Verified &0.234 & 0.250 & 0.007  \\
$\sigma$(User partisanship) & 0.098 & 0.003 & 0.021 \\
  $\mu$|User partisanship) & 0.028 &  0.005 & 0.007 \\
  $\mu$(User Toxicity) &  0.589 & 0.486 & 0.502 \\
    \bottomrule

    \end{tabularx}
  \caption{Pearson correlation $\rho$ and Kendall's $\tau$ of dependent variables and clusters' toxicities. } 
   \vspace{-15pt}
   \label{table:regression-clusterlevel}
\end{table}

As seen in Table~\ref{table:regression-clusterlevel}, and Figure~\ref{fig:partial-dependcies-clusterlevel}, unsurprisingly, the most important factor in determining the toxicity of a given topic is the toxicity of the users contributing tweets to the cluster. This one variable has a permutation importance of 0.50 and a correlation of 0.58 with the toxicity of a given cluster. Simply put, unsurprisingly, topics whose corresponding users have higher average toxicity are more likely to have toxic content. As in Section~\ref{sec:toxic_middle}, we again observe being further along the political spectrum does not necessarily indicate increased toxicity and that a conversation being dominated by right-leaning or left-leaning users has little bearing on its toxicity. 

We find, as seen in Figure~\ref{fig:partial-dependcies-clusterlevel}, that the number of users involved in a given topic appears to have a moderating and mitigating effect on the toxicity of that topic ($\rho=0.292$). This also appears as one of the most important features for determining the average toxicity with a permutation score of 0.445. However, conversely having more verified individuals participating in that topic \emph{does} increase toxicity. We thus find (from Section~\ref{sec:toxic_middle}) that while verified users are less likely to tweet toxic content, their presence and their tweeting about particular topics correlate with increased toxicity in that topic. We further find that despite the average age of accounts participating in a topic having a negative Pearson correlation in our fitted model if the average age of the accounts participating in a conversation is very young or much older, there is a decreased toxicity compared to topics that engage accounts of all ages.

Examining political ideological contributions in Figure~\ref{fig:partial-dependcies-clusterlevel}  to the toxicity of individual topics we find that topics dominated by all left-leaning or all right-leaning users are largely the least topics compared to topics in the middle of the ideological spectrum. Finally examining the partial dependence of the diversity of viewpoints that participate in a given topic at a given point in time, we find that while initially the greater the political diversity of the topic cluster, the more toxic it becomes, as the topic invites more and more users of different beliefs that the topic cluster decreases in toxicity. While further research is needed, this result reinforces the work of Mamkos et~al. that find that for particular typically non-political topics that engage users from all over the political spectrum, these topics tend to be less toxic than others~\cite{mamakos2023social}.

 \begin{figure}
\begin{minipage}[l]{0.6\textwidth}
\includegraphics[width=1\columnwidth]{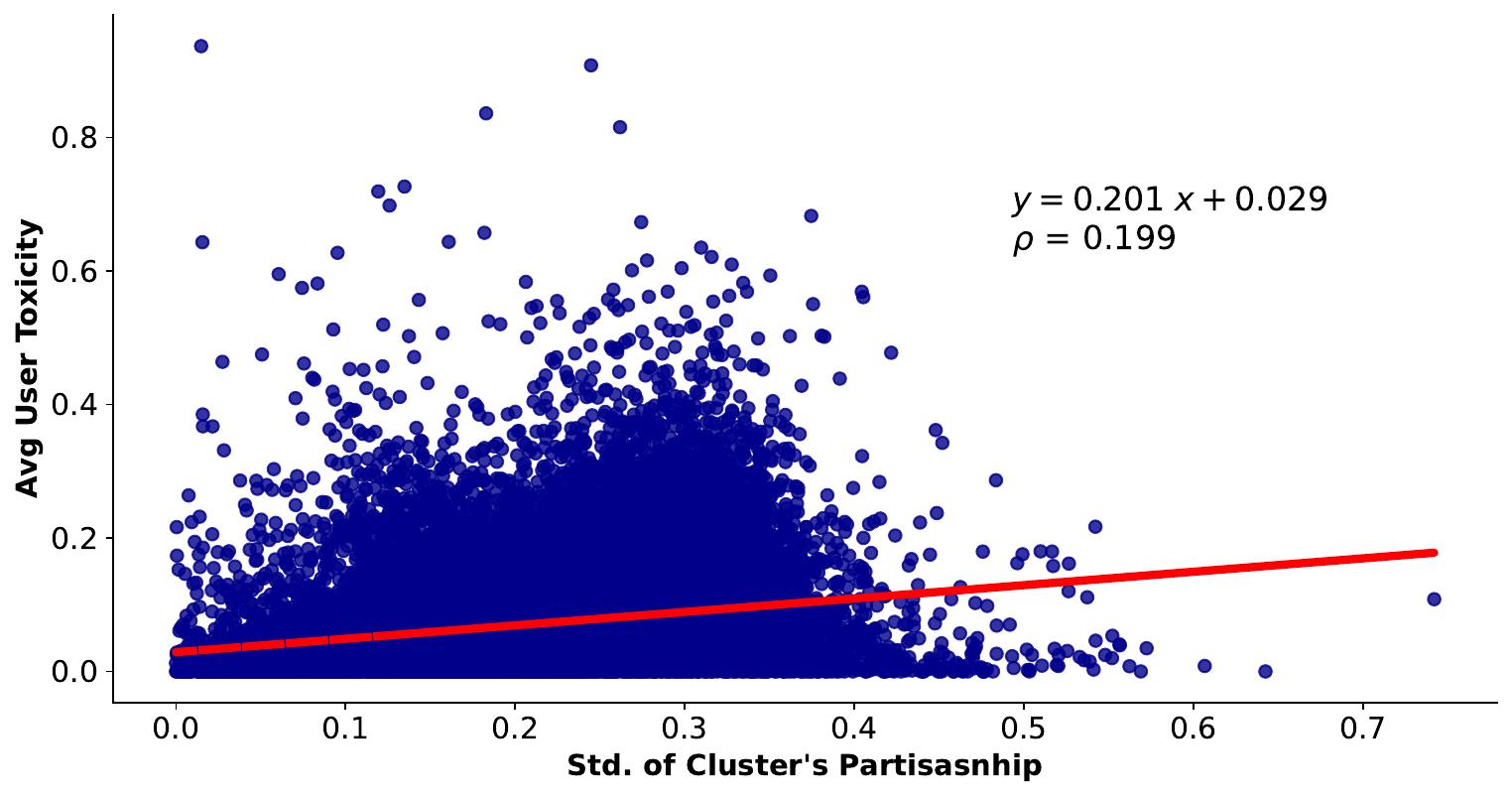} 
\end{minipage}
\begin{minipage}[l]{0.33\textwidth}
\caption{As previously also found in our analysis of user characteristics (Section~\ref{sec:calculated}), we find that as users engage in a wider window of topics of particular political ideologies, the more toxic their tweeted content\label{fig:toxicity-vs-var-user}}
\end{minipage}

\end{figure}

Lastly, looking in the reverse direction, we determine how users' toxicity changes when they are involved in many different types of politically aligned topics.  As also found by Mamkos et~al.~\cite{mamakos2023social}, we find, as seen in Figure~\ref{fig:toxicity-vs-var-user},  as users are involved in a higher variance of topics of different political orientations their average toxicity increases ($\rho=0.19$). We thus find from this analysis further confirmation, on a topic level, that increased user toxicity and the diversity of views present in a given conversation contribute to toxicity within particular topics. However, conversely, as topics invite a wider range of individuals into a discussion toxicity actually decreases. We now consider some of the implications of these results.

%


\section{Discussion}
In this work, we determined the correlation of different aspects of partisanship and affective polarization with toxicity at a user and topic-level on Twitter. We find, most notably, that users who are at the tail end of the political spectrum (very right-leaning or very left-leaning) \emph{are not} more likely to post toxic content; rather, we observe that users that engage with a wide variety of different politically aligned accounts center have a higher likelihood of tweeting toxic messages. Further, as users interact with or mention other users from a wider range of political ideologies, they are more likely to post toxic content. We similarly find that users who interact with other users who more regularly post toxic content are more likely to post toxic content themselves. 

Examining these phenomena from a topic level, we find that most heavily partisan topics \emph{are not} the most toxic. Rather, topics often have complex relationships with the partisanship of the users who tweet about them. While some topics become more toxic as more right-leaning/left-leaning users tweet about them, others become less toxic. However, as with individual users, we find that as users from a wider range of political ideologies tweet about a given topic, the more toxic that topic. Here we discuss some of the limitations and implications of our results:

\subsection{Limitations}
We acknowledge several limitations of this work. Given our use of GAMs to estimate the effect of partisanship and political diversity and our lack of ability to perform direct experiments, our findings are largely correlational. While they do buttress and support a large literature of similar results~\cite{barbera2014social,barbera2015tweeting,dagoula2019mapping,bail2018exposure} that have found causal results in some cases for increased polarization due to interaction with users of different political beliefs, we acknowledge that \emph{our} results are not causal. We further note that due to new restrictions placed on the collection of Tweets~\cite{Singh2023}, we can not continue to measure the toxicity of users and political topics, going forward.

This work largely focuses on US-based political polarization and ideologies. As a result, while applicable to dynamics for Twitter accounts on the US-political spectrum, our results do not necessarily apply to political conversations in different contexts. However, given access to Twitter or similar social media websites such as Meta's Threads, our study can be replicated in different cultural contexts.

Finally, as found early in our work in Section~\ref{sec:labeling-toxic}, different individuals and datasets have different metrics for toxicity. While our use of Perspective API's definition of toxicity is standard throughout the literature~\cite{kumar2021designing,rajadesingan2020quick,saveski2021structure}, we do base our DeBERTa-based model toxicity detection on this definition; we acknowledge that it may not take into account all perspectives on what constitutes toxic online content.

\subsection{Tribal Tendencies, Affective Polarization, Online Toxicity, and Online Echo Chambers on Twitter}
As found by others, heated political conversations often elicit toxicity as people of differing views debate and discuss their differences~\cite{salminen2020topic}. We find that this discourse is related to increased toxicity on Twitter. The political diversity of those involved in a given Twitter conversation surrounding a given topic, at least in the short form of tweets, is correlated with affective polarization and toxic content. Adding nuance to previous studies of communities that have found that like-minded users gather and reinforce each other's views, creating toxic echo-chambers~\cite{cinelli2020echo,starbird2018ecosystem,gronlund2015does}. While users naturally often congregate and more heavily engage with users like themselves (assortativity coefficient of 0.266), showing that some echo chambers may exist on Twitter, when users exit these chambers and engage with other users of differing political views, we observe that this tends to create user conflict~\cite{guess2018avoiding}. This result reinforces De Francisci Morales {et~al.}'s~\cite{de2021no} finding that interactions among users on Reddit with different political orientations have increased negative conversational outcomes, showing that it occurs in platform-wide user interactions and discussions. Further indeed across all users, we find that as they increasingly interact with users of different partisanship, the frequency of toxicity increases (Figure~\ref{fig:toxicity_vs_mention-new}). While this feature of online conversation is not the dominant factor in engendering toxic content, with other factors like a user's previous behavior~\cite{levy2022understanding}, the age of their account, and the toxicity of other users also contributing to online toxicity, we note that this apparent ``tribal tendency'' appears both on a user and topic-level across Twitter and across multiple Twitter threads illustrating the robustness of this finding~\cite{mamakos2023social,de2021no}.  


\subsection{Hyperpartisan Users and Topics}
In contrast to some prior work~\cite{nelimarkka2018social},  we find that users and topics that are hyperpartisan (\textit{i.e.}, very left-leaning users or very right-leaning users) are not necessarily more toxic than less ideological users. Rather, we find these users tend to mostly associate and interact with other users who share similar political views ($\rho=0.605$) and as a result, do not necessarily have higher toxicity levels. As also found by Gr{\"o}nlund et al.~\cite{gronlund2015does}, because hyperpartisan users and topics often do not attract users of differing political views, we find that these users and topics tend to be less toxic than topics and users that interact with a wider range of the political spectrum (\textit{e.g.}, topics and users nearer to the political center). This result indicates that political echo chambers, where only left-leaning or right-leaning interact among themselves may be less conflict-oriented on Twitter. As a result, we argue that if social media companies like Twitter wish to expose their users to a wider range of political views without increasing conflict on their platforms, these users may be amenable to these opposing views if they come from others nearer to themselves on the political spectrum.


\subsection{Intra-Topic Partisanship over Time}
In Section~\ref{sec:changes}, we observed that the political orientation of users who discuss any particular topic often changes over time. These changes, often coinciding with changes in toxicity, also illustrate that the views expressed on Twitter about particular topics often change as different users enter or leave conversations. We argue that future analysis of topics and their spread on Twitter \emph{must} take into account user-level characteristics such as partisanship given that these values often reveal the nature of how users are addressing individual topics. For example, as seen in Section~\ref{sec:changes}, understanding that conversations surrounding ``Moscow Mitch'' had been taken up by increasingly right-leaning users reveals the penetration of this insult into more conservative circles. 


\subsection{Toxic Birds of Feather}
In addition to finding that the range of political views encountered by a particular user is predictive of toxicity, we further find that topics and users who interact with other toxic users are more likely to be toxic themselves. This again buttresses prior work from Kim {et~al.}, Kwon {et~al.}, and Shen {et~al.}  who all find that exposure to these negative conversations actually increases observers' tendency to also engage in incivility~\cite{kim2019incivility,kwon2017offensive,shen2020viral}. While not a new finding~\cite{kumar2023understanding}, this illustrates that reducing toxic content online may have other downstream benefits; by removing more instances of toxic content, other users may be less likely to engage in toxicity themselves further reducing the amount of toxic content. Given the existence of particular toxicity norms within communities Reddit~\cite{rajadesingan2020quick}, where toxicity is rarely seen among users and toxic comments are looked down upon, we argue that removing toxic content may have a compounding effect, greatly improving the overall health of online discourse.

\subsection{Implications for the Twitter/X Platform} Our work simultaneously finds that topics that engage with a wider set of politically aligned users and that users that engage in a wider array of different political discussions are more likely to tweet toxic messages. Namely, exposure on the Twitter/X platform to differing views may essentially be counterproductive to producing civil online discussions~\cite{bail2018exposure}.  Furthermore, this suggests that recent attempts to widen the range of political discussion on Twitter may have the additional effect of increasing online toxicity~\cite{hickey2023auditing}. As such, we argue that as Twitter continues to widen the political conversation on its platform, to also maintain low levels of toxicity additional moderation steps or additional practices should be taken to slowly introduce users to other accounts with different political beliefs to themselves should be taken as well~\cite{munson2010presenting}. This accords with the recommendations and findings of Mamakos et al.~\cite{mamakos2023social} who found that as Reddit users engage with users different from them and in a wider variety of political contexts, they tend to be more toxic. Given that Twitter users are not siphoned out into individual communities that they specifically join and thus more easily engage with polarizing content and users with whom they disagree across their topics of interest, we argue that building a means by which to engage in better conversations across political differences can reduce toxicity and friction on the platform. For example, as also argued by~\cite{nelimarkka2018social} including a wide and generalized view of particular topics could potentially reduce polarization. Indeed, as found in Section~\ref{sec:topic-level-gam}, while initially sparking more toxicity, as topics include a wider and wider berth of political perspectives and as more users join a topic, the toxicity of that particular topic decreases. 



\subsection{Future Work}
This work centered around understanding factors that contribute to the toxicity levels of individual users and within particular topics on Twitter/X. However, we note that several of the techniques employed within this work can be extended and utilized beyond our study.


\vspace{2pt}
\noindent
\textbf{Identifying the Role of Partisanship and Polarization on Different Platforms} In this work, while we focus on Twitter, we note that our approach can largely be utilized on different social media platforms (\textit{e.g.}, Facebook, Reddit, \textit{etc...}) to identify the role of partisanship and political polarization. Unlike on Twitter, where a feed is curated for the user, Reddit user interactions, for instance, are largely determined by the into which the communities self-select. Previous work has shown that entire communities can engage in cross-partisan toxic behavior~\cite{efstratiou2022non}. Similarly Bail et~al.~\cite{bail2018exposure} find that simply following users and repeatedly seeing disagreeable content can increase polarization. As such, we plan to explore the robustness of our findings about ``tribal tendencies'' in different contexts and what best practices can be utilized to ameliorate these tendencies.

\section{Conclusion}
In this work, we analyzed what factors potentially contribute to 
why users post toxic content on Twitter. We propose and implement a new open-source toxicity classifier, achieving better accuracy than the Perspective API on the Civil Comments dataset.  Then, analyzing 89.6M tweets posted by 43.1K users from across the political spectrum, we find a user or topic being heavily partisan does not necessarily imply increased toxicity; rather as users engage with and as conversations involve a wider range of political orientations and with other toxic users and toxic content that online toxicity increases.


\bibliographystyle{ACM-Reference-Format}
\bibliography{sample-base}

\appendix
\section{Correspondence Analysis for Approximating Political Ideology\label{sec:appendix-ca}}
After identifying our set of 882 politically discriminating and identifying 7,7 random accounts that followed this set of accounts, we performed the following for CA. 

\begin{enumerate}
\item{\textbf{Identify the Ideological Subspace:}} Using 6,107 accounts that followed 10 or more of our 882 discriminating political users, we derive an initial CA model and obtain a discriminating latent space on which to plot user political ideology. 

\item{\textbf{Expand the number of discriminating political ideological accounts:}} Utilizing our initial CA model we determine the set of Twitter accounts not included within our initial target accounts that were most often followed by the most conservative and liberal accounts (within the top 20\% on either side of the political spectrum) in the first stage of our analysis. As in Barbera et~al.~\cite{barbera2015tweeting}, we compute the popularity among users of a given ideological orientation such that $pop_{jc} = n_{jc} -  n_{jl}$  for conservatives, where $n_{jc}$ is the number of conservative users included in the first stage
that follow account j, and $n_{jl}$ is the equivalent measure for liberals. We further filter these accounts to ensure that at least 3r different users follow these additional discriminating accounts. After determining these users, we add the resulting 788 accounts as additional "following" accounts to our original $n \times m$ matrix. These additional accounts include those of Barack Obama (@BarackObama), MSNBC (@MSNBC), Florida governor Ron Desantis (@GovRonDeSantis), and the House GOP (@HouseGOP).

\item{\textbf{Expanding the number of follower accounts:}} For the rest of our users, we project them into the discriminating latent space utilizing our CA model. This allows us to utilize the information from our original discriminating political accounts as well as from the additional discriminating political accounts from the second stage. We further can estimate the political ideology of any account that follows at least one of 1607 highly politically discriminating accounts. After projecting all of our users we standardize the estimates into z-scores  (\textit{i.e.,} a value of 0 represents the average partisanship and a value of 1 represents one standard deviation above the mean, 2, two standard deviations above the mean, \textit{etc...}). Altogether we project an additional 49,308 users. 


\end{enumerate}

\section{Unsupervised Contrastive Learning\label{sec:finetune}}
We utilize the SimCSE training objective to further refine our MPNet model and ensure that it is properly suited for our dataset. This is such that we embed each tweet $i$  $x_i = (tweet_i) \in D_{tweets}$ (where $tweet_i$ is the text) twice (with dropout both times) using MPNet by inputting $[CLS] text_i [SEP]$ and outputting out the contextual hidden vectors $\mathbf{h}_i$ and $\tilde{\mathbf{h}}_i$  for $text_i$ as its representations. Then, given a batch of contextual hidden vectors $\{\mathbf{h}_i\}_{i=0}^{N_b}$ and $\{\tilde{\mathbf{h}}_{j}\}_{j=0}^{N_b}$ (different dropout), where $N_b$ is the size of the batch, for each batch in our training dataset of 1 million tweets, we perform a contrastive learning step on that batch. This is such that for each batch $\mathcal{B}$, for an \textit{anchor} hidden embedding $\mathbf{h_i}$ within the batch, the set of hidden contextual vectors $\mathbf{h_i} \, \mathbf{\tilde{h_j}} \in \mathcal{B}$, 
 the hidden contextual vectors where $i = j$ are positive pairs. Other pairs where $i\neq j$ are considered negative pairs. Within each batch $\mathcal{B}$, the contrastive loss is computed across all positive pairs in the batch such that:
\[
    L_{contrastive} = -\frac{1}{N_b} \sum_{\mathbf{h}_i \in \mathcal{B}}\mathit {l}^c(\mathbf{h}_i )
\]
\[
\mathit{l}^c(\mathbf{h}_i) = {log}\frac{ \sum_{j\in\mathcal{B} } \mathbbm{1}_{[i = j]}\mathrm{exp}( \frac{\mathbf{h}_i^\top \tilde{\mathbf{h}_j}}{\tau||\mathbf{h}_i || ||\tilde{\mathbf{h}_j} || })}{\sum_{j\in\mathcal{B}} \, \mathrm{exp}( \frac{\mathbf{h}_i^\top \tilde{\mathbf{h}_j}}{\tau||\mathbf{h}_i || ||\tilde{\mathbf{h}_j} || })}
\]
where, as in prior work~\cite{liang2022jointcl}, we utilize a temperature $\tau=0.07$.

\section{Training our Open-Source Toxicity Classifier\label{sec:app-tox-classifier}}

\vspace{2pt}\noindent
\noindent
\textbf{Realistic Adversarial Perturbations of the Civil Comments Training Dataset.} To train our model, we rely on the Civil Comments training dataset which consists of 1,804,874~comments that were each individually graded by up to 10~human raters for their toxicity. Each comment, depending on the percentage of human raters that graded the comment as ``toxic'' (toxic having the definition provided in Section~\ref{sec:misinformation-defintion}), is assigned a score between 0 and 1. Our  training dataset is thus $D_{Civil} = \{x_i = (text_i,t_i)\}^N_{i=1}$, where $text_i$ a text, and $t_i$ is the toxicity of the text. While the Civil Comments training dataset is fairly large, we note that it is heavily skewed with 1,268,269 of the comments having a toxicity score of 0. To ensure that our training dataset has a wider set of examples of comments with above zero estimated toxicity, we augment the Civil Comments training dataset using realistic adversarial perturbations~\cite{le2022perturbations}.

Utilizing the ANTHRO dataset provided by Le  {et~al.}~\cite{le2022perturbations}, for every comment with above zero toxicity within the Civil Comments dataset, we leverage the set of common human-written perturbations to augment our Civil Comments dataset. This ANTHRO dataset consists of common online perturbations of words (\textit{e.g.}, Republican $\rightarrow$ republiican, Reeepublican, Republicaan) extracted from online texts (\textit{e.g.}, Twitter). For each comment with a toxicity score greater than zero in the Civil Comments training set, we extract a set of random perturbations of each noun and adjective within the comment, perturbing the overall comment nine times with different combinations of the perturbed nouns and adjectives. This enables us to extend the set of non-zero comments to a total of 5,366,050~comments (6,634,319 in the full augmented dataset). We utilize this dataset when training our DeBERTa-based~\cite{he2022debertav3} model to determine the toxicity of tweets. We note that in addition to allowing our model to have more training instances of toxic texts, this approach further enables our model to have training instances of real ``in-the-wild'' perturbations and misspellings of words that are often found on social media (\textit{e.g.}, Twitter) and online.

\vspace{2pt}\noindent
\noindent
\textbf{DeBERTa-based Contrastive Embedding Layer.} Besides utilizing our augmented dataset of realistic adversarial perturbations, while training our model, we pre-train a contrastive layer to differentiate toxic and non-toxic texts. We later freeze this layer while training our full model to identify the toxicity of individual tweets.

To pre-train this layer for use in our model, we utilize contrastive learning to differentiate toxic and non-toxic texts. As in the original Civil Comments task, while training this layer we consider texts with labeled toxicity $t_i$ $>0.5$ score in the Civil Comments dataset as toxic and those with labeled toxicity $t_i$ $<0.5$ as nontoxic. We utilize this threshold for classifying a comment as toxic, given that this score (as described in the Civil Comments task) indicates that a majority of the Civil Comments annotators would have assigned a ``toxic'' attribute to this comment. For training, this is such that we embed each example $x_i = (text_i,t_i) \in D_{Civil_{aug}}$ (where $text_i$ is the text and $t_i$ is whether the text is toxic or not) using a contextual word model by inputting $[CLS] text_i [SEP]$ and outputting the hidden vector $\mathbf{h}_i$ of the [CLS] token for each $text_i$ as its representation. Then, given a set of hidden vectors $\{\mathbf{h}_i\}_{i=0}^{N_b}$, where $N_b$ is the size of the batch, we perform a contrastive learning step on that batch. This is such that for each Batch $\mathcal{B}$, for an \textit{anchor} hidden embedding $\mathbf{h_i}$ within the batch, the set of hidden vectors $\mathbf{h_i} \,, \mathbf{h_j} \in \mathcal{B}$ vectors where $i \neq j$, we consider them a positive pair if $t_i, t_j$ are equivalent. Other pairs where $t_i\neq t_j$ are considered negative pairs. Within each batch $\mathcal{B}$, the contrastive loss is computed across all positive pairs in the batch such that:

\[
    L_{toxic} = -\frac{1}{N_b} \sum_{\mathbf{h}_i \in \mathcal{B}}\mathit {l}^c(\mathbf{h}_i )
\]
\[
\mathit{l}^c(\mathbf{h}_i) = {log}\frac{ \sum_{j\in\mathcal{B}\setminus i } \mathbbm{1}_{[t_i = t_j]}\mathrm{exp}( \frac{\mathbf{h}_i^\top \mathbf{h}_j}{\tau||\mathbf{h}_i || ||\mathbf{h}_j || })}{\sum_{j\in\mathcal{B}\setminus i } \, \mathrm{exp}( \frac{\mathbf{h}_i^\top \mathbf{h}_j}{\tau||\mathbf{h}_i || ||\mathbf{h}_j || })}
\]
\begin{figure}
\begin{minipage}[l]{0.5\textwidth}
\includegraphics[width=1\columnwidth]{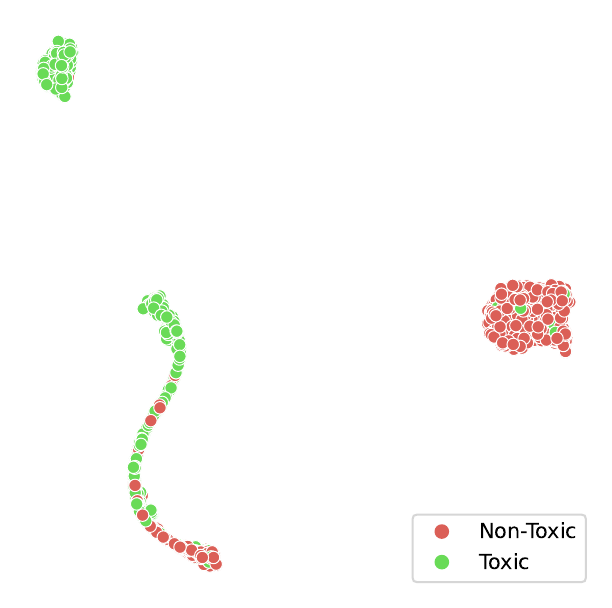} 
\end{minipage}
\begin{minipage}[l]{0.35\textwidth}
\caption{\textbf{t-SNE of Civil Comments Validation Dataset }-- As we train the DeBERTa-based contrastive embedding layer of our model on our augmented Civil Commonents training set, our model can differentiate non-toxic (\textit{i.e.,} toxicity $t_i < 0.5$) and (\textit{i.e.,} toxic $t_i > 0.5$) comments. However, comments that are of ambiguous toxicity are more difficult to differentiate.\label{figure:tsne} }
\end{minipage}

\end{figure}

\noindent
where, as in prior work~\cite{liang2022jointcl}, we utilize a temperature $\tau=0.07$. Throughout training, we use a batch size of 64 and a learning rate of $1\times 10^{-5}$, training for three epochs. After training this layer, we freeze it for use in the rest of our model. As seen in Figure~\ref{figure:tsne}, reducing the dimensionality of the outputted $h_{constrat}$ on the Civil Comments validation dataset using t-SNE~\cite{van2008visualizing}, our contrastive embeddings are largely though imperfectly, able to differentiate between non-toxic and toxic comments.

\begin{figure}
\begin{minipage}[l]{0.5\textwidth}
\includegraphics[width=1\columnwidth]{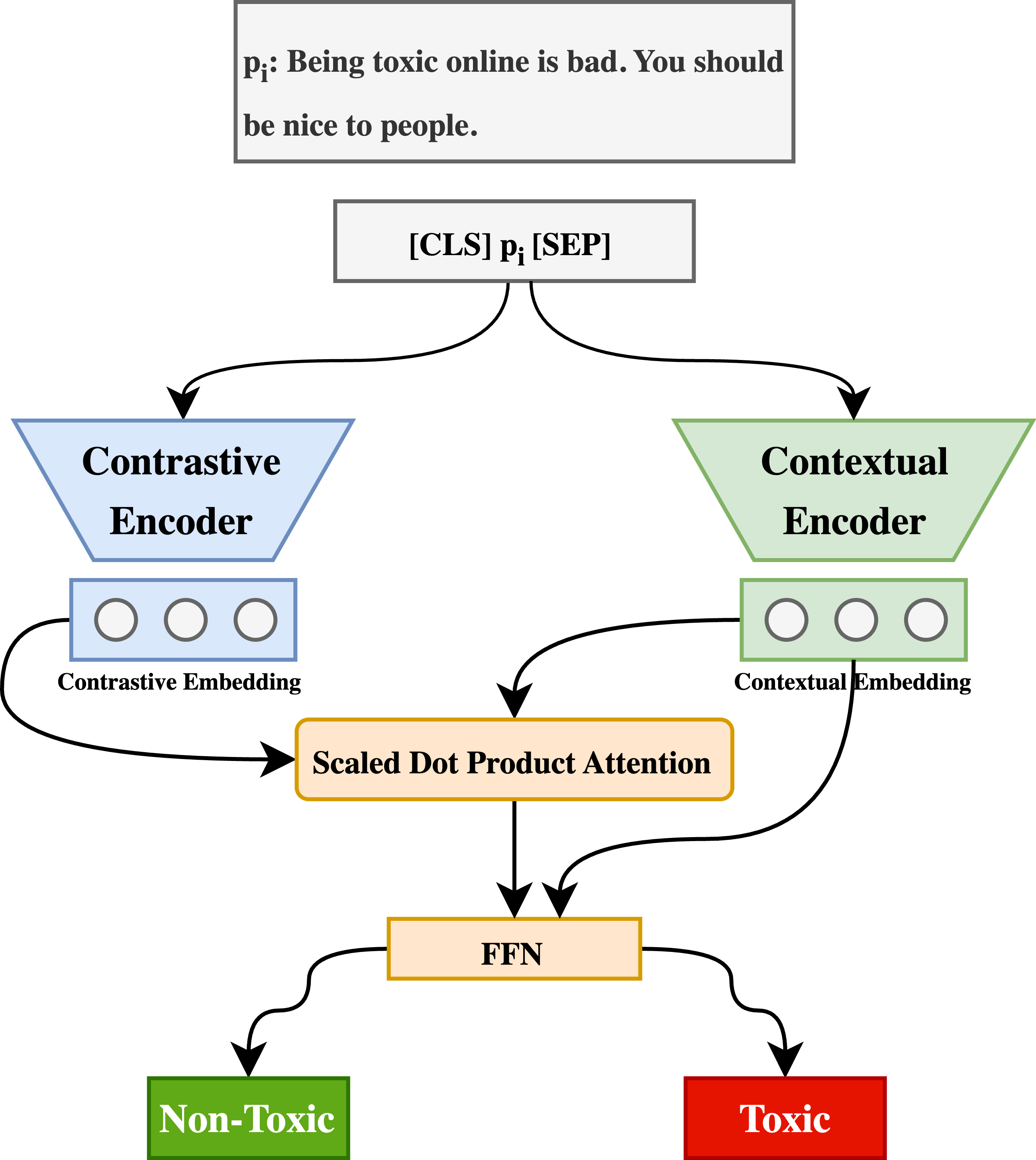} 
\end{minipage}
\begin{minipage}[l]{0.35\textwidth}
\caption{{Model to determine the toxicity of individual tweets}--- We utilize contrastive learning, scaled-dot-product attention, and the DeBERTa model to train a model to predict the toxicity of tweets in our dataset. Our fully trained model achieves a 0.818 Pearson correlation with the toxicity scores in the Civil Comments test dataset. \label{fig:toxicity-twitter-model}}
\end{minipage}

\end{figure}

\vspace{2pt}\noindent
\noindent
\textbf{Full DeBERTa Toxicity Detection Model.} Taking our pretrained-DeBERTa contrastive embedding layer and our augmented dataset $D_{Civil_{aug}}$, we finally train our full DeBERTa toxicity detection model (Figure~\ref{fig:toxicity-twitter-model}. This model first computes the scaled dot product of a DeBERTa hidden representation of a text $h_{text}$ and the $h_{contrast}$ output of our DeBERTa contrastive embedding layer. The intuition behind this approach is to enable our model to determine the extent of the toxicity features present within the original text.  

\begin{align*}
r_{contrast} &= \sum_i a_ih_{text}^{(i)},\\
a_i &= \textrm{softmax} \left( \lambda h_{text}^{(i)} \cdot (W_{contrast} h_{contrast}) \right)
\end{align*}

\noindent
where and $\lambda = 1/\sqrt{E}$, $E=$ dimentionality of the the embeddings, and $W_{contrast}$ is a learned parameter matrix. Finally, once $r_{contrast}$ is calculated, we concatenate it using a residual connection with the original $h_{text}$. We then feed the resulting representation into a feed-forward network with ReLU activation for determining the toxicity of the text as seen in Figure~\ref{fig:toxicity-twitter-model}. We minimize mean squared error while training, utilizing the Civil Comments validation dataset to perform early stopping with a patience of 2. Throughout training, we use a batch size of 64 and a learning rate of $1\times 10^{-5}$. We completed all training on a Nvidia A6000 GPU\@. 

\section{Pointwise Mutual Information\label{sec:pmi} }

The pointwise mutual information PMI of a particular word $word_i$ in a cluster $C_j$ is calculated as:

\vspace{-10pt}
\begin{align*}
\scriptsize
PMI(word_i, C_j) = log_2\frac{P(word_i,C_j)}{P(word_i) P(c_i)}
\end{align*}
\vspace{-10pt}

\noindent where $P$ is the probability of occurrence and a scaling parameter $\alpha$ is added to the counts of each word. This scaling parameter $\alpha$ prevents single-count or one-off words in each cluster from having the highest PMI values. Given the scale of our dataset and the number of clusters within our dataset, we determine that a baseline count of 1 ($\alpha$ =1) for each word in the full dictionary in each cluster led to the best results~\cite{turney2001mining}.

\section{DP-Means\label{sec:ap-dpmeans}}

DP-Means~\cite{kulis2011revisiting} is a non-parametric extension of the K-means algorithm that does not require the specification of the number of clusters \textit{a priori}. Within DP-Means, when a given datapoint is a chosen parameter $\lambda$ away from the closest cluster, a new cluster is formed. Dinari {et~al.}~\cite{dinari2022revisiting} parallelize this algorithm by \textit{delaying cluster creation} until the end of the assignment step. Namely, instead of creating a new cluster each time a new datapoint is discovered, the algorithm determines which datapoint is furthest from the current set of clusters and then creates a new cluster with that datapoint. By delaying cluster creation, the DP-means algorithm can be trivially parallelized. Furthermore, by delaying cluster creation, this version of DP-Means avoids over-clustering the data (\textit{i.e.,} only the most disparate data points create new clusters)~\cite{dinari2022revisiting}.

\clearpage
\newpage
\section{GAM Fit of of User-Level Features and Perspective Toxicity\label{sec:perspective-user-app}}
\begin{figure}[!h]
\begin{minipage}[l]{1.0\textwidth}
\includegraphics[width=1\columnwidth]{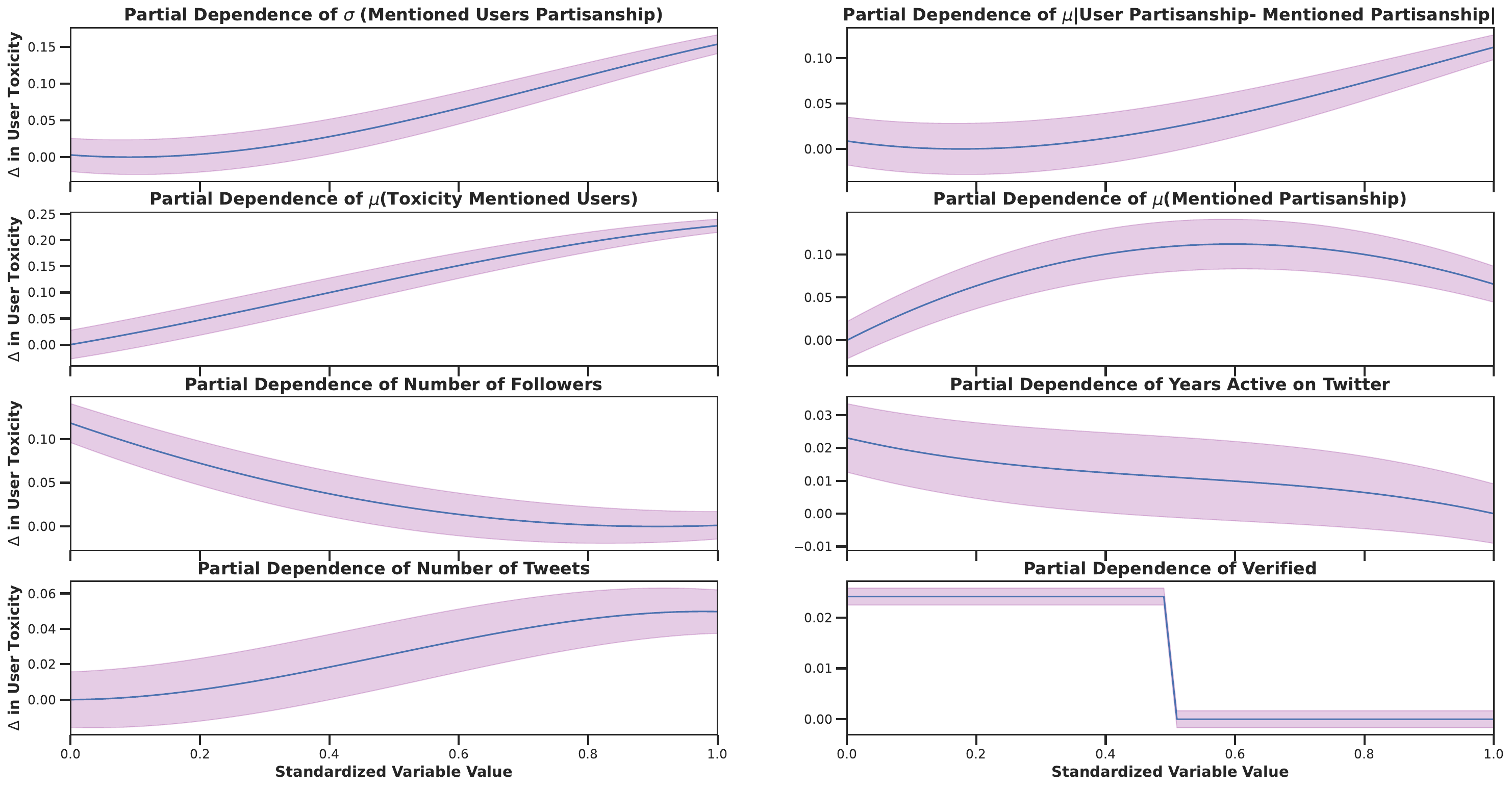} 
\end{minipage}
\begin{minipage}[l]{1\textwidth}
\caption{Partial dependencies with 95\% Normal Confidence intervals between fitted standardized dependent variables and user Perspective API toxicity.}
\end{minipage}

\end{figure}

\begin{table}[!h]
    \small
    \centering
    \begin{tabularx}{0.9\columnwidth}{l|rrr}
    Train $R^2$ 0.266, Validation $R^2$:  0.270 \\
    \toprule
      Dependent Variable  & Pearson Corr. $\rho$ & Kendall's $\tau$ & Permut Import. \\    \midrule
  Verified Status & ---- &-0.233& 0.031  \\
  Years Active on Twitter & -0.220& -0.155 & 0.022 \\
  Log \# Followers & -0.229 &-0.137 & 0.231  \\
  Log \# Followed & -0.197 & -0.128 & --- \\
  Log \# Tweets in 2022 & 0.182 & 0.173  & 0.094\\
  Toxicity of Mentioned Users& 0.366& 0.347 & 0.409 \\
  Partisanship  &0.075& 0.079 & --- \\
  $\sigma$(Mentioned Users Partisanship)&0.331& 0.294	& 0.149\\
  $\mu$|User Partisanship- Mentioned Partisanship|&0.272& 0.241 & 0.015\\
  $\mu$(Mentioned Partisanship)& 0.139 &0.114 & 0.048  \\
    \bottomrule
    \end{tabularx}
  \caption{Pearson correlation $\rho$ and Kendall's $\tau$ of dependent variables and user's individual toxicity. As seen in the above table, a user's interaction with a wide political variety of users and interacting with other users with higher toxicity correlates with a given user's own toxicity. } 
   \vspace{-15pt}
   \label{table:regression-not-important}
\end{table}

\clearpage
\newpage

\section{The Most Partisan Toxic Topics of 2022\label{sec:partisan-toxic}}
In this section, we give an overview of the most partisan topics in 2022.

\begin{table*}
\centering
\scriptsize
\selectfont
\setlength{\tabcolsep}{4pt}
\begin{tabularx}{\textwidth}{lXrrrXrrr}
\toprule
 &   &  &   \# Toxic & Avg. & Example & Avg.  & Avg. Partisan  & Partisan \\
 Topic& {Keywords}&\# Tweets & Tweets & Toxicity&  Tweet  & Partisan. & of Toxic Users& Std. \\

\midrule
1 & nra/russia, dispelled, finance, thoroughly, parroting &105 & 99 (94.29\%) &0.767 & The NRA/Russia narrative was proven to be complete bullshit by the Senate Finance Committee investigation. The Democrats came off looking like imbeciles. Now you look like an imbecile for parroting a thoroughly dispelled narrative &  2.550 &2.712 & 0.839\\
2 & tock, tick, cleaned, clock, 22, november & 18,780 & 171 (0.91\%) &0.077 &  727 days until the next election on Tuesday, November 5, 2024. Start working now to take the oval office, the senate the house. PS: Brandon's son didn't die in Iraq; he's a sexual and incestuous pervert;the worst president in US history. & 2.024 & 2.024& 0.000\\
3 & chemtrail, nanoparticle, poisoning, 31, murdering & 69 & 64 (92.75\%)  &0.718& YOU KNOW OF THE TRUMP BIDEN MINISTRY OF SATAN NAZI WORLD WAR 2 HOLOCAUST CHEMTRAIL GENOCIDE POISONING TECHNOLOGY USED BY TRUMP AND BIDEN BUT DO NOTHING! & 1.456 &  -0.166& 0.626\\

4 &bannons, rustyrockets, joerogan, planet, ingraham & 608& 90 (14.8\%) &0.140 & Disgusting and horrific! Reminiscent of Nazi Germany! Could put political opponents in here!? Outrageous! Fox News Maria Bartiromo Bret Baier marthamaccallum Bannons War Room Prison Planet & 1.396 &0.727 &0.258\\

5 & eagle, patriot, red, railfan, 1,187 &1,587 & 100 (8.42\%)&0.0948& As opposed to "establishment favorite" which is utter bs, you should say crowd interested in actually being able to win with someone. & 1.311 &0.950 &1.113\\

\bottomrule
\end{tabularx}
\caption{\label{tab:conservative-topics} Top toxic topics by right-leaning tilt in our dataset.} 
\end{table*}

\begin{table*}
\centering
\scriptsize
\selectfont
\setlength{\tabcolsep}{4pt}
\begin{tabularx}{\textwidth}{l|XrrrXrrr}
\toprule
 &   &  &   \# Toxic  & Avg.& Example & Avg.  & Avg. Partisan  & Partisan \\
 Topic& {Keywords}&\# Tweets & Tweets & Toxicity &  Tweet  & Partisan. & of Toxic Users& Std. \\

\midrule
 1 & marjorie, pardon, greene, nazi, traitorous &239 & 69 (28.97\%) &0.312 & You're one of the "others" YOU SEDITIOUS TRAITOR 
 & -2.903 &-1.180& 1.363\\

  2 & mastriano, thanmaga, antisemitic, louder, mastribator & 128& 74 (57.82\%) &0.492 & Sen Mastriano You're an anti-Semitic POS.
 & -2.176& -0.682 & 1.791 \\
  3 & conor, lamb, pa, ahaha, stans &1,619& 54 (3.33\%)&0.057 & Wow...You guys are Pathetic. Conor Lamb has PLENTY of Grassroots support. I am one of them. Conor is also Endorsed by the Majority of Unions. So Factsmatter PA Sen Lamb for US senate!! 
 & -1.796 &-0.774 & 1.245\\

 4 &alleged, attacked, treason, above, smeared &1,311& 212 (16.17\%) &0.239 & 
This is our Nations no 1 problem
The lies that come from here
Tearing up our society
Backing Trump who destroyed our Country let in Russians to our House. & -1.784& -0.651 & 0.432\\

 5 & livable, kyrsten, centrist, survive, sinema & 160& 125 (78.13\%)&0.547 & Many people don't want us to survive or to have a livable planet because to them rich people's bank accounts matter more! Looking At Republicans Joe Manchin Kyrsten Sinema& -1.604& -0.739 &0.652\\
\bottomrule 
\end{tabularx}
\caption{\label{tab:liberal-topics} Top toxic topics by left-leaning tilt.} 
\vspace{-10pt}
\end{table*}


\vspace{2pt}
\noindent
\textbf{Right-Leaning Topics.} As seen in Table~\ref{tab:conservative-topics}, we observe that the most right-leaning topic concerned animus towards the media and US government for alleging the US National Rifle Association was an asset of the Russian government and spread Russian propaganda during the 2016 election~\cite{Mack2019}. Beyond, this topic, we further a tweet reminding Republican voters of the date of the next presidential election while simultaneously calling President Joe Biden the worst president in history and his son a pervert (Topic 2). We further find a series of tweets about the ``Chemtrails conspiracy theory'' alleging that the US government is killing its citizens~\cite{xiao2021sensemaking}. Finally, we observe several tweets angry at the establishment and the US government (Topics 4 and 5), with users calling US policies ``reminiscent of NAZI Germany.''

\vspace{2pt}
\noindent
\textbf{Left-Leaning Topics.}
As seen in Tables~\ref{tab:liberal-topics}, in several cases, many of the most left-leaning topics simply disparage right-leaning political figures (Topic 1, 2, 4 in Table~\ref{tab:liberal-topics}). These targets include current US Republican public officials and candidates like the Georgia Congresswoman Marjorie Taylor Greene~\cite{Donnelly2022}, former US President Trump (Topic 4), and Pennsylvania gubernatorial candidate Greg Mastriano. Beyond these three officials, we further observe attacks against Independent Senator Arizona Kyrsten Sinema and Democratic West Virginian Senator Joe Machin for rebelling against Democratic leadership in the Senate~\cite{Teh2022}. We lastly observe in Topic~3, many left-meaning accounts defending former Pennsylvanian Congressman Connor Lamb when he ran in the Democratic primary for an open Senate seat~\cite{Zipkin2022}.

\begin{figure}
\begin{minipage}[l]{0.45\textwidth}
\includegraphics[width=1\columnwidth]{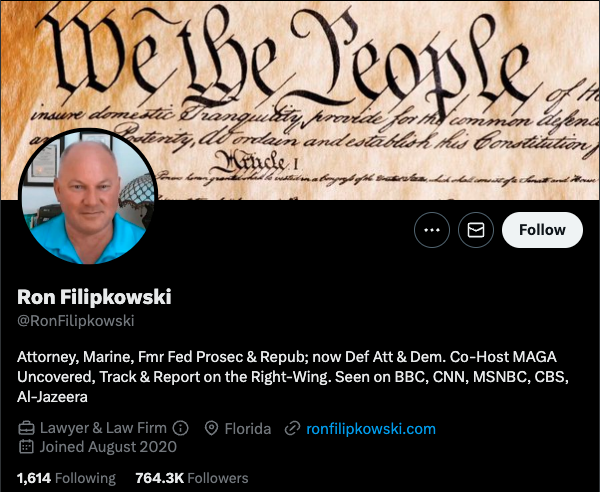}
\end{minipage}
\begin{minipage}[l]{0.45\textwidth}
\includegraphics[width=1\columnwidth]{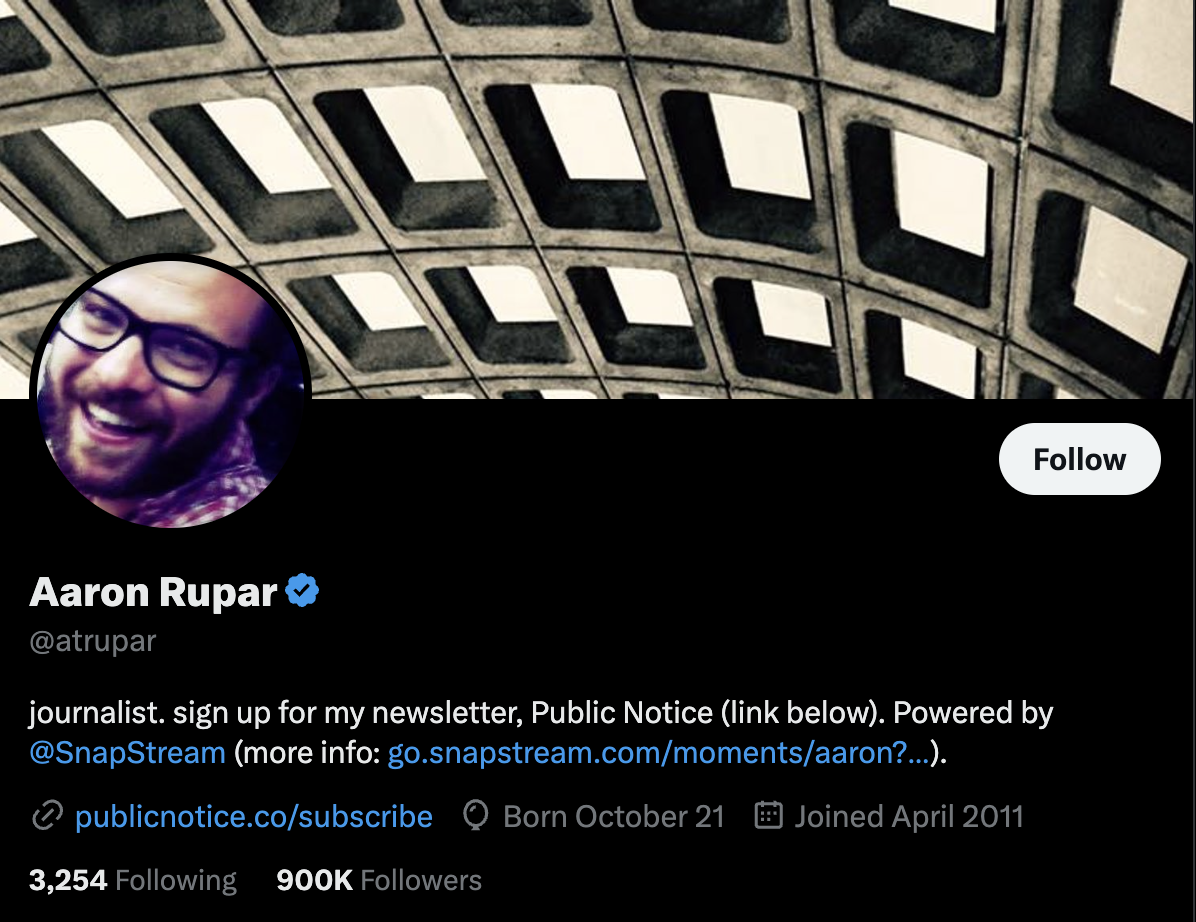} 
\end{minipage}

\begin{minipage}[l]{1\textwidth}
\caption{Accounts like @ronfilipkowski and @atrupar had users of similar political orientations reply in a toxic manner to the news and opinions they tweeted.\label{fig:orientation-bio}}
\end{minipage}

\end{figure}

\begin{table}
\centering
\small
\begin{tabular}{l|rr}
\textbf{Account} &\textbf{\# Waves } & \textbf{Avg Partisanship of Wave } \\\midrule
@ronfilipkowski & 25  & -0.491\\
@youtube & 18 & 0.128\\
@elonmusk & 16 & 0.345 \\
@potus & 16 & 0.537\\
@repmtg & 14 & -0.366\\
@laurenboebert & 12 & 0.091\\
@acyn & 11& -0.766\\
@atrupar & 10 &-0.583\\
@donaldjtrumpjr & 10 & -0.131\\
@foxnews & 9 & -0.090\\
\bottomrule

\end{tabular}
\caption{\label{tab:campaigns} Number and average partisanship of toxic reply/mention waves encountered by various Twitter accounts.} 
\vspace{-15pt}
\end{table}
\vspace{2pt}
\noindent
\textbf{``Toxic Topics Waves.''} As was seen in our set of left-leaning topics, after examining the rest of our clusters, we observed several other separate instances when various users encountered ``toxic topics waves'' of toxic replies/mentions (\textit{i.e.}, where the majority of tweets were "@"s at a particular account).  For example, while we displayed one toxicity wave targeting Georgia Congresswoman Majorie Taylor Greene, we identified 13 other such toxicity waves in our dataset. For example, in one such wave, a user wrote:
\begin{displayquote}
\small
\textit{Oh wow RepMTG all the traitors together today!!!!  Could y'all imagine if the Obamas or Clinton's did this corrupt bullshit!!}
\end{displayquote}
while in another wave a different user wrote
\begin{displayquote}
\small
\textit{RepMTG Marge is a neanderthal idiot. No one is stupidly pushing drag queen shows or teaching gender lies - advocating "genital mutilation" ffs - what a bunch of ridiculous stupid! But hey, JAN 6 coup to sell the US to Russia, DANGER of losing our Democracy to extremists wacks!}
\end{displayquote}

Altogether we identified 4,506~toxicity waves against 3,822~users. 1,383 of these waves have a right-leaning orientation (\textit{i.e.}, average partisanship of toxicity wave participant > 0) while 3,123 have a liberal orientation. Calculating the political orientation of these ``attacked'' accounts, across these ``toxicity waves'', 14.5\% were in cases of right-leaning accounts campaigning against liberal accounts; 17.4\% were cases of liberal accounts campaigning against right-leaning accounts; 35.9\% were right-leaning against right-leaning; 33.5\% were left-leaning against left-leaning. Compared to all mentions where only 33.8\% are between users of different political orientations, we thus again observe evidence of affective polarization in these ``toxic topics waves.'' In Table~\ref{tab:campaigns}, we present the number of ``topics toxicity waves'' against particular users. 54.5\% (541 accounts) of the ``attacked'' accounts were verified (compared to only 10.7\% [4,610 accounts] of the accounts out dataset of 43,151 Twitter users),  suggesting that more public figures are more likely to incur these waves.

We note that, while in some cases these are targeted campaigns meant to attack particular users, in several cases these toxic waves are other Twitter accounts toxicly responding in agreement to the opinions or news put forward by the account. While the waves targeting @laurenboebert, a conservative congressperson from Colorado, are mostly by heavily left-leaning users for example, this occurs in reverse for two hyperpartisan liberal commentators user @atrupar and  @ronfilipkowski. For example in one such case, a user tweeted
\begin{displayquote}
\small
\textit{@RonFilipkowski Trump supporters are so dumb, they confuse antifa with nazis. There were nazis in Trump's White House.}
\end{displayquote}
Other waves, for instance, were aimed at @YouTube to protest particular videos being taken down.  


\clearpage
\newpage
\section{Most Toxic Topics \label{sec:most-toxic-by-percentage}}
\begin{table}[!h]
\centering
\scriptsize
\selectfont
\setlength{\tabcolsep}{4pt}
\begin{tabularx}{\textwidth}{l|XrXrXrrr}
\toprule
 &   &  &   \# Toxic & Avg.& Example & Avg.  & Avg. Partisan  & Partisan \\
 Topic& {Keywords}&\# Tweets &Tweets  & Toxicity &  Tweet  & Partisan. & of Toxic Users& Std. \\
\midrule

 1 & fuck, shit…, shit, shittttt, extremely &52 & 52 (100\%) &0.923 & That's all folks. Fuck this shit. &-0.169 &-0.167 & 0.737\\ 
2 & idiot, blitering, complete, total, he & 3121 & 3,123 (99.94\%) &0.915 & Not idiots. Deliberate enablers of fascism. &  0.152 & 0.128 & 0.970 \\
3& fuck, you, him, though, that's & 340 & 336 (98.82\%) &0.902& Fuck this and fuck him.& -0.027& -0.117 & 0.836 \\
4& piece, load, shit, hahahha, you &756 & 775 (97.55\%) &0.895 &   Tell me you are a piece of shit without telling me. & 0.011 &0.007& 0.957 \\
5&volume, youtube, chop, stupid, that & 435 & 438 (99.32\%) &0.880 & Nothing stupid about that!  & 0.119&0.104& 0.942 \\

\bottomrule
\end{tabularx}
\caption{\label{tab:narratives} Top toxic topics---by average toxic value---in our dataset.\label{table:toxic-topic-max}} 
\end{table}

\newpage
\section{Linear Fit of Topic-Level Features against Perspective Toxicity \label{sec:cluster-perspective-tox}}
\begin{figure}[!h]
\begin{minipage}[l]{1.0\textwidth}
\includegraphics[width=1\columnwidth]{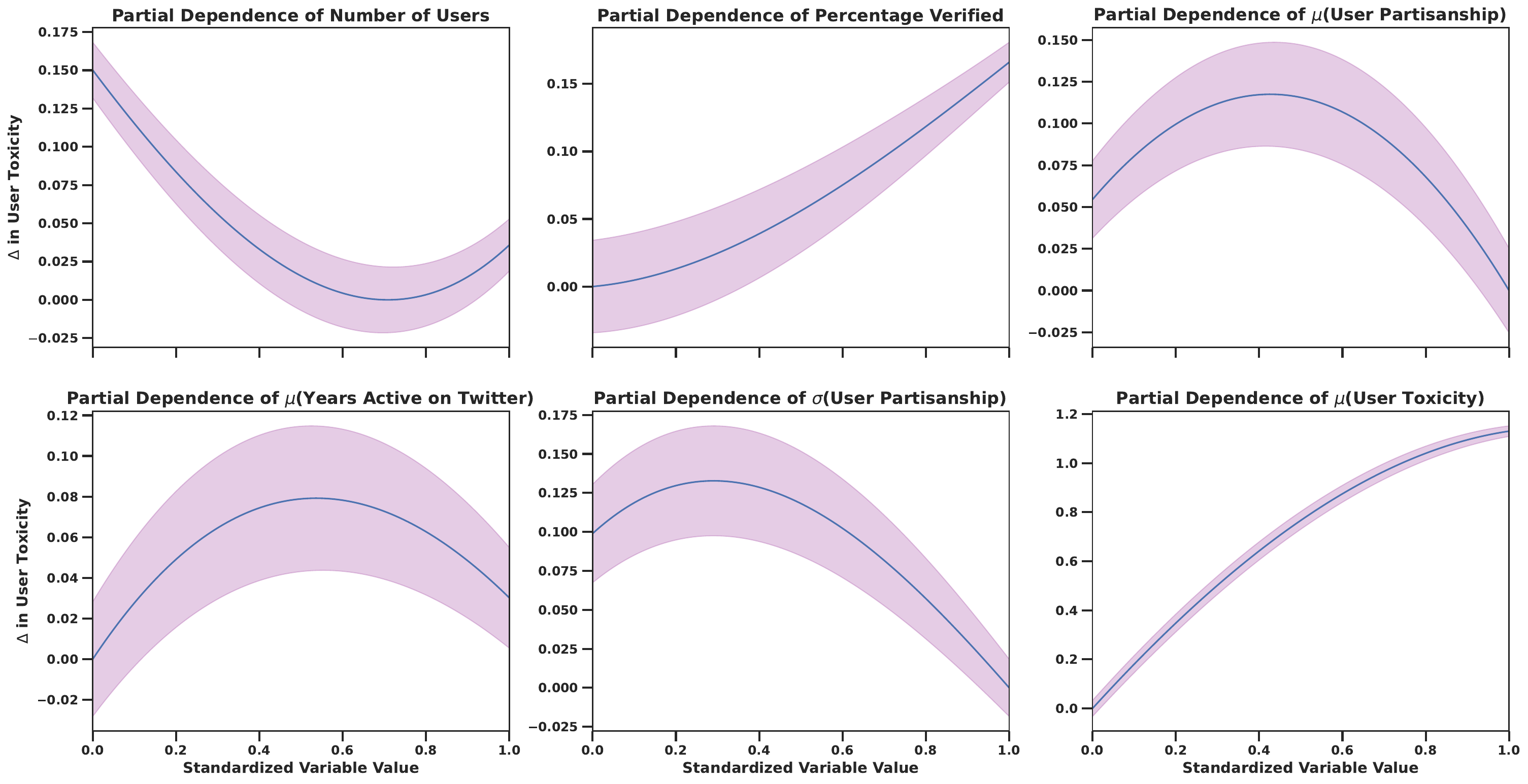} 
\end{minipage}
\begin{minipage}[l]{1\textwidth}
\caption{Partial dependencies with 95\% Normal Confidence intervals between fitted standardized dependent variables and cluster Perspective API toxicity.}
\end{minipage}
\end{figure}
\begin{table}[!h]
    \small
    \centering
    \begin{tabularx}{0.85\columnwidth}{l|rrr}
      Train $R^2$ 0.454, Validation $R^2$:  0.463  \\
    \toprule
      Dependent Variable  & Pearson Corr. $\rho$ & Kendall's $\tau$ & Permut Import. \\    \midrule
  Number of Users &-0.268 & -0.132  & 0.520  \\
 $\mu$(Years Active on Twitter) &-0.233 & -0.192 & 0.010\\
  Percentage Verified &0.273 & 0.247 & 0.014 \\
$\sigma$(User Partisanship) & -0.097 & -0.012 & 0.036 \\
  $\mu$|User Partisanship) & -0.014 &  0.011 & 0.013 \\
  $\mu$(User Toxicity) &  0.637 & 0.502 & 0.398 \\
    \bottomrule
    \end{tabularx}
  \caption{Pearson correlation $\rho$ and Kendall's $\tau$ of dependent variables and clusters' toxicities. } 
   \vspace{-15pt}
   \label{table:regression-20240402}
\end{table}

\end{document}